\documentclass[journal]{IEEEtran}
\IEEEoverridecommandlockouts
\usepackage{booktabs}
\usepackage{siunitx}
\usepackage{cite}
\usepackage{amsmath,amssymb,amsfonts}
\usepackage{graphics, framed}
\usepackage{graphicx, enumerate}
\usepackage{epsfig}
\usepackage{epstopdf}
\usepackage{url}
\usepackage{multimedia}
\usepackage{paralist}
\usepackage[printonlyused]{acronym}
\usepackage{units}
\usepackage{algorithmic}
\usepackage{textcomp}
\usepackage{float}
\usepackage{tabularx}
\usepackage{balance}
\usepackage{subfigure}
\usepackage{xcolor}
\def\BibTeX{{\rm B\kern-.05em{\sc i\kern-.025em b}\kern-.08em
		T\kern-.1667em\lower.7ex\hbox{E}\kern-.125emX}}
\usepackage{mathtools}
\usepackage{cuted} 
\usepackage[export]{adjustbox}

\usepackage{amsmath}
\usepackage{amssymb}
\usepackage{dblfloatfix}
\usepackage{lipsum}
\usepackage{color,soul}
\usepackage{multirow}
\usepackage[overload]{textcase}
\newcommand{\FGR}[1]{Fig.~\ref{#1}}

\newcommand{\SEC}[1]{Section~\ref{#1}}
\newcommand{\TAB}[1]{Table~\ref{#1}}

\newcommand{\minus}{\scalebox{0.75}[1.0]{$-$}}
\acrodef{5G}[5G]{5\textsuperscript{th}-Generation}
\acrodef{BW}[BW]{bandwidth}
\acrodef{BER}[BER]{bit error rate}
\acrodef{BPSK}[BPSK]{binary phase-shift keying}
\acrodef{CW}[CW]{continuous wave}
\acrodef{CSI}[CSI]{channel state information}
\acrodef{D2D}[D2D]{device-to-device}
\acrodef{dB}[dB]{decibel}
\acrodef{dBi}[dBi]{decibel isotropic}
\acrodef{dBm}[dBm]{decibel over a milliwatt}
\acrodef{DSN}[DSN]{deep space network}
\acrodef{DTN}[DTN]{delay tolerant network}
\acrodef{Gbps}[Gbps]{gigabit per second}
\acrodef{GHz}[GHz]{gigahertz}
\acrodef{GAT}[GAT]{graph attention network}
\acrodef{THz}[THz]{Terahertz}
\acrodef{ISL}[ISL]{inter-satellite link}
\acrodef{RIS}[RIS]{reconfigurable intelligent surface}
\acrodef{GM}[GM]{Gamma mixture}
\acrodef{PSK}[PSK]{phase shift keying}
\acrodef{QAM}[QAM]{quadrature amplitude modulation}
\acrodef{AWGN}[AWGN]{additive white Gaussian noise}
\acrodef{SNR}[SNR]{signal-to-noise ratio}
\acrodef{AF}[AF]{amplitude-and-forward}
\acrodef{MIMO}[MIMO]{multiple-input multiple-output}
\acrodef{mMIMO}[mMIMO]{massive-multiple-input multiple-output}
\acrodef{SDN}[SDN]{Software-defined network}
\acrodef{SON}[SON]{self-organizing network}
\acrodef{HetNet}[HetNet]{heterogeneous network}
\acrodef{FSO}[FSO]{free-space optics}
\acrodef{UM-MIMO}[UM-MIMO]{ultra-massive-MIMO}
\acrodef{AP}[AP]{access point}
\acrodef{UE}[UE]{user equipment}
\acrodef{NTN}[NTN]{non-terrestrial network}
\acrodef{UAV}[UAV]{unmanned aerial vehicle}
\acrodef{HAPS}[HAPS]{high-altitude platform station}
\acrodef{LEO}[LEO]{low-Earth orbit}
\acrodef{BAN}[BAN]{body area network}
\acrodef{WLAN}[WLAN]{wireless local area network}
\acrodef{QoS}[QoS]{quality of service}
\acrodef{TCS}[TCS]{thermal control system}
\acrodef{QCL}[QCL]{quantum cascade laser}
\acrodef{CMOS}[CMOS]{complementary metal-oxide semiconductor}
\acrodef{V-HetNet}[V-HetNet]{vertical heterogeneous network}
\acrodef{DL}[DL]{Deep learning}
\acrodef{DRL}[DRL]{deep reinforcement learning}
\acrodef{EIRP}[EIRP]{effective isotropic radiated power}
\acrodef{FDTD}[FDTD]{Finite-difference time-domain}
\acrodef{FEM}[FEM]{finite element method}
\acrodef{MoM}[MoM]{method of moments}
\acrodef{VNA}[VNA]{vector network analyzer}
\acrodef{CS}[CS]{channel sounder}
\acrodef{CIR}[CIR]{channel impulse response}
\acrodef{CTF}[CTF]{channel transfer function}
\acrodef{DPM}[DPM]{Dirichlet process mixture}
\acrodef{TOA}[TOA]{time of arrival}
\acrodef{GMM}[GMM]{Gaussian mixture model}
\acrodef{IoT}[IoT]{Internet of Things}
\acrodef{MLE}[MLE]{maximum likelihood estimation}
\acrodef{LOS}[LOS]{line-of-sight}
\acrodef{NLOS}[NLOS]{non-line-of-sight}
\acrodef{SG}[SG]{signal generator}
\acrodef{SEP}[SEP]{Sun-Earth-probe}
\acrodef{FDSOI}[FDSOI]{fully depleted silicon on insulator}
\acrodef{OpEx}[OpEx]{operational expenditures}
\acrodef{TCO}[TCO]{total cost of ownership}
\acrodef{CapEx}[CapEx]{capital expenditures}
\acrodef{MAC}[MAC]{medium access control}
\acrodef{GEO}[GEO]{geostationary orbit}
\acrodef{SWaP}[SWaP]{size, weight, and power}
\acrodef{NOMA}[NOMA]{Non-orthogonal multiple access}

\ifCLASSINFOpdf
\else
\fi

\renewcommand{\baselinestretch}{0.987}

\begin{document}
	\title{Energy-Efficient RIS-Assisted Satellites for \\IoT Networks}

	\author{K{\"{u}}r{\c{s}}at~Tekb{\i}y{\i}k,~\IEEEmembership{Graduate Student Member,~IEEE,} G{\"{u}}ne{\c{s}}~Karabulut~Kurt,~\IEEEmembership{Senior~Member,~IEEE,} Halim~Yanikomeroglu,~\IEEEmembership{Fellow,~IEEE}

		\thanks{K. Tekb{\i}y{\i}k and G.K. Kurt are with the Department of Electronics and Communications Engineering, {\.{I}}stanbul Technical University, {\.{I}}stanbul, Turkey, e-mails: \{tekbiyik, gkurt\}@itu.edu.tr}
		\thanks{G. Karabulut Kurt is also with the Department of Electrical Engineering, Polytechnique Montr\'eal, Montr\'eal, Canada, e-mail: gunes.kurt@polymtl.ca}
		
		\thanks{H. Yanikomeroglu is with the Department of Systems and Computer Engineering, Carleton University, Ottawa, Canada, e-mail: halim@sce.carleton.ca}
		
	}
	
	\IEEEoverridecommandlockouts 
	
	\maketitle

	\begin{abstract}
		The use of satellites to provide ubiquitous coverage and connectivity for densely deployed \ac{IoT} networks \textcolor{ black}{is} expected to be a reality in emerging 6G networks. Yet the low battery capacity of \ac{IoT} nodes constitutes a problem for \textcolor{ black}{their} direct connectivity to satellites\textcolor{ black}{, which} are located \textcolor{ black}{at altitudes of up to 2000 km.} \textcolor{ black}{In this paper, we propose a novel architecture involving the use of} \ac{RIS} units to \textcolor{ black}{mitigate} the path loss associated with \textcolor{ black}{long} transmission distances. \textcolor{ black}{These \ac{RIS} units can be placed on satellite reflectarrays, and, when used in broadcasting and beamforming, they can provide significant gains in signal transmission.} \textcolor{ black}{This study shows} that \ac{RIS}-assisted satellites can provide up to $10^5$ times higher downlink and achievable uplink rates for \ac{IoT} networks.
	\end{abstract}
	
	\IEEEpeerreviewmaketitle
	\acresetall

	\section{Introduction}\label{sec:intro}
	\textcolor{ black}{Vendors and business advisory companies expect that around 100 billion devices will be connected in a massive ecosystem by 2025}~\cite{whitepap86:online, TheInter67:online}. \ac{IoT} networks are expected to grow \textcolor{ black}{by} approximately 20 percent compound annual growth rates~\cite{DIIoTPri87:online}. By leveraging ubiquitous \ac{IoT} networks, it is possible to \textcolor{ black}{increase} the efficiency in industries such as transportation, health, and maritime. Another prominent feature of massive \ac{IoT} networks is \textcolor{ black}{that they improve} quality of life. Although ubiquitous and massive \ac{IoT} networks \textcolor{ black}{have a broad range of advantages for industry and human life}, backhaul of these ultra-massive networks \textcolor{ black}{is still challenging issue}. \textcolor{ black}{Furthermore, Internet access is still limited in some parts of the world. For instance, 4 billion people do not have an Internet connection}~\cite{kota2019satellite}. \textcolor{ black}{To expand coverage area}, supporting IoT services with satellite networks is a prominent research topic for both academia and industry.
	
	\begin{figure*}[ht]
		\centering
		\includegraphics[width=\linewidth, page=1]{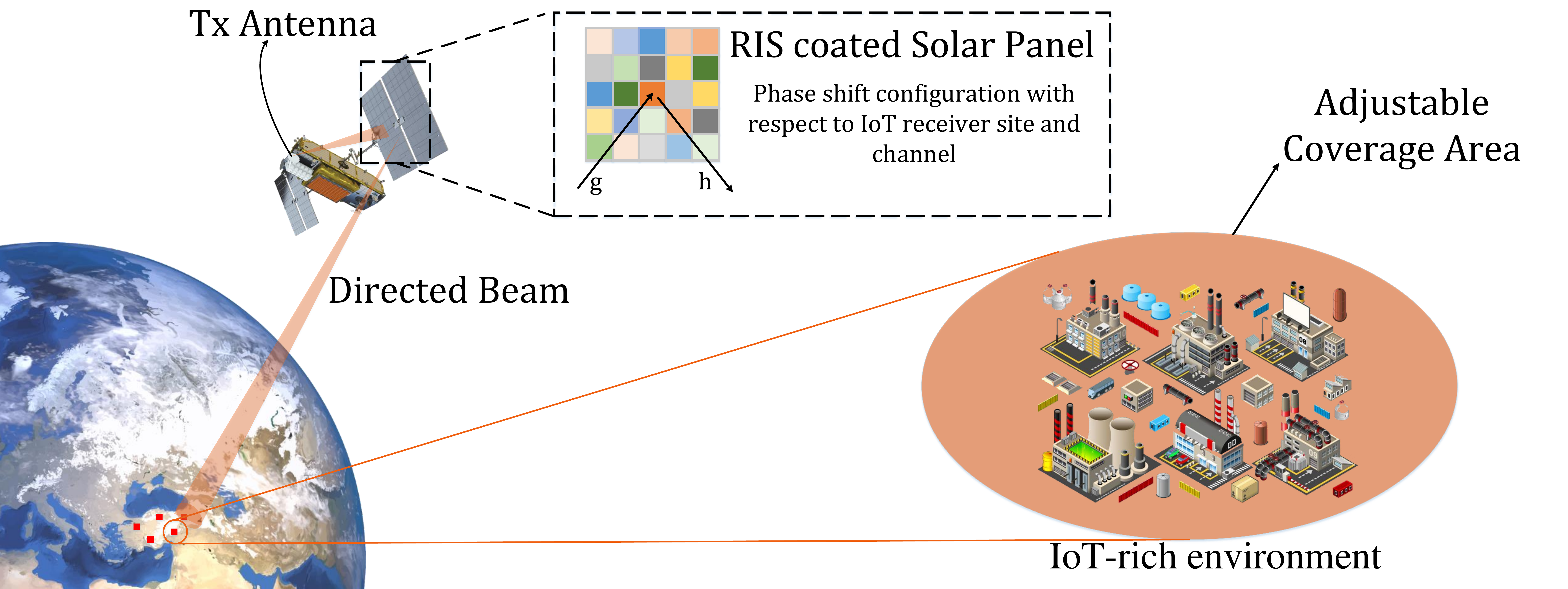}
		\caption{\textcolor{ black}{It is possible to enhance the QoS for satellite-IoT systems by utilizing RISs. Therefore, the required power can be reduced for the same data rate and error probability.}}
		\label{fig:system_scheme}
	\end{figure*}

	\subsection{Related Works}\label{sec:related_works}
	\textcolor{ black}{Several recent works have} proposed for the use of \ac{GEO} satellites in narrowband-\ac{IoT} applications~\cite{hofmann2019direct, barbau2020nb}. However, \ac{GEO} satellites \textcolor{ black}{suffer high path losses and long delays due to their considerable distance from the Earth.} To eliminate the drawback of \ac{GEO} satellites, \ac{LEO} satellite constellations \textcolor{ black}{, which orbit the Earth closer than GEO satellites, were introduced for \ac{IoT} networks in}~\cite{qu2017leo, cluzel20183gpp}. \ac{LEO} satellites require less transmission power \textcolor{ black}{to maintain a requisite \ac{SNR} for proper communication, but they have drawbacks as well. In particular, the high velocities of LEO satellites make the steering antennas and satellite tracking systems essential.} \textcolor{ black}{Because \ac{IoT} devices have low battery capacity and low computing power, they cannot track \ac{LEO} satellites and run high-complexity signal processing algorithms in the receiver unit. Thus, IoT devices need low complexity transceivers and energy-efficient signal processing and tracking methods.} In order to \textcolor{ black}{track satellites and decrease computational cost at receiver in an energy-efficient way, we propose using} \acp{RIS} in \ac{LEO} satellites in this study. \acp{RIS} intelligently adjust the phase shifts of elements in order to maximize the received power~\cite{basar2019wireless}. \textcolor{ black}{Furthermore, RISs are also advantageous because they consume less power and have lower hardware costs than conventional arrays. This is because conventional arrays have bulky feeding networks, which result in significant power losses. By contrast, RIS phase control mechanisms use simple PIN diodes~\cite{dai2020reconfigurable}.} \textcolor{ black}{But,} the most appealing feature of \acp{RIS} is that \textcolor{ black}{they are entirely comprised of passive circuit elements and require no complex processing or coding.} \textcolor{ black}{The effectiveness of RISs has been demonstrated by experimental measurements in}~\cite{tang2019wireless, dai2020reconfigurable}. \textcolor{ black}{Metasurface technologies, which include RISs, have also recently been proposed for deep space communications in \cite{gonzalez2020metasurface, yurduseven2020towards, wang2020metantenna, rotshild2019wideband}. By utilizing \acp{RIS}, it is possible to improve performance in terms of power consumption and/or diversity gain.}

	\textcolor{ black}{The installation of RIS units on LEO satellites does not} require serious hardware improvements, as \acp{RIS} have a simple hardware structure consisting of passive circuit elements~\cite{tekbiyik2020reconfigurable}. Moreover, the hardware complexity of \ac{IoT} devices can be reduced as \acp{RIS} allow signal processing to be performed in the transmission environment rather than on the devices~\cite{basar2019wireless}. Thus, battery-constrained \ac{IoT} devices can operate for a longer duration. Furthermore, \acp{RIS} \textcolor{ black}{can} enable energy-efficient wireless communication while \textcolor{ black}{maintaining the same} \ac{QoS} level~\cite{wu2019intelligent}. Our previous work~\cite{tekbiyik2020reconfigurable} \textcolor{ black}{has also} demonstrated that error probability can be reduced in \ac{LEO} inter-satellite links by utilizing \acp{RIS}. 
	
	\subsection{Motivations}\label{sec:motivations}
	\textcolor{ black}{As the number of IoT applications and devices increases exponentially, the need for ubiquitous Internet coverage is becoming increasingly apparent. LEO satellites offer a promising solution for this.} \textcolor{ black}{With a high number of satellites in near-Earth constellations, it will become possible to ubiquitously and continuously service users.} \textcolor{ black}{The provision of IoT services via} LEO networks has been under investigation not only by the research community but also by standards organizations, such as 3GPP, and private companies, such as Satelliot. Especially with the 5G standards, following the 3GPP TR 36.763~\cite{3GPP36.763}, IoT deployments \textcolor{ black}{involving} satellites will be a reality in the near future, not only \textcolor{ black}{for} rural areas that do not get sufficient coverage but also \textcolor{ black}{for} improving the capacity in highly populated dense deployments in mega-cities. 
	
	As \ac{IoT} devices have limited battery \textcolor{ black}{life}, any boost in the link budget will be useful in satellite networks. To address this issue in this study\textcolor{ black}{,} we propose \textcolor{ black}{the use of RISs on the receiver end to improve received power levels}. \ac{RIS} units, which are also called meta-atom, are composed of metasurfaces that are controlled in real time to adjust the reflection phases~\cite{basar2019wireless}. \textcolor{ black}{The potential of RISs} has already been \textcolor{ black}{discussed} in recent works\textcolor{ black}{,} including~\cite{huang2019reconfigurable, wu2019intelligent}. In this study, we propose the use of \ac{RIS} \textcolor{ black}{on} satellites that can be positioned jointly with a reflective array as illustrated in \FGR{fig:system_scheme}. \textcolor{ black}{Due to the large surface area under satellite solar panels, \textcolor{ black}{numerous RIS units can be installed.} By configuring the phase shift for each meta-atom, the incident wave can be scattered or beamformed \textcolor{ black}{to} target users. Furthermore, by properly selecting phase shifts, the coverage area can be adaptively adjusted with respect to the stochastic geometry of \ac{IoT} networks. In investigating the use of RISs in satellite IoT systems, we show that it is possible to reduce power consumption and hardware costs of IoT devices. This is made possible by the fact that RISs can increase the received \ac{SNR} and reduce computational complexity by eliminating the need for complex signal processing applications at the receiver end.}
	
	\vspace{-0.3cm}
	\subsection{Contributions}\label{sec:contributions}
	\textcolor{ black}{The contributions of the paper} on the way to provide ubiquitous and dense connectivity demanded by 6G and beyond are summarized \textcolor{ black}{here}:
	\begin{enumerate}[{C}1]
		\item A novel architecture is proposed for \ac{IoT} networks based on the use of \ac{RIS} units \textcolor{ black}{with \ac{LEO}} satellites. These are \ac{RIS} units can be placed under the solar panels or replaced with reflect arrays.
		\item \textcolor{ black}{The SNR levels of LEO satellites for supporting IoT networks is quantified, taking into consideration broadcasting and beamforming modes based on transmission characteristics, including carrier frequency.}
		\item Through extensive numerical results, \textcolor{ black}{we quantify the potential reduction of the transmit power using a realistic transmission model}. Based on our analysis, we suggest design guidelines for future \ac{IoT} networks served by LEO satellites.
	\end{enumerate}

	\subsection{Outline}\label{sec:outline}
	
	The rest of this paper is organized as follows. \SEC{sec:preliminaries} introduces the free\textcolor{ black}{-}space path loss models for \ac{RIS}-assisted broadcasting and \ac{RIS}-assisted beamforming schemes as well as \textcolor{ black}{a} case without \ac{RIS}. In \SEC{sec:leo_system_model}, the system model is described for \ac{RIS}-assisted \ac{LEO} satellites for \ac{IoT} networks. \SEC{sec:results} presents extensive numerical results and observations for non-\ac{RIS} satellites and \ac{RIS}-assisted satellites with both broadcasting and beamforming schemes. Open issues are discussed in \SEC{sec:open_issues}. Finally, the study is concluded in \SEC{sec:conclusion}.
	
	\section{Preliminaries}\label{sec:preliminaries}
	\textcolor{ black}{Here} we discuss the mathematical and physical preliminaries for \ac{RIS}-assisted satellites for \ac{IoT} networks. \textcolor{ black}{We begin with an overview of free-space path loss for conventional satellite systems. Then, we introduce a path loss model for RIS-assisted wireless communications.} \textcolor{ black}{In what follows}, the notation is given for \textcolor{ black}{the} downlink; however, \textcolor{ black}{the notation} can be generalized for \textcolor{ black}{the} uplink case in a straightforward manner.
	
	\subsection{Free-Space Path Loss}
	
	The ratio of the received and transmitted powers in a link between two isotropic antennas can be given by free-space path loss, which is defined as follows:
	\begin{align}
	L_{\mathrm{FS}}=(4 \pi d / \lambda)^{\alpha},
	\label{eq:pl}
	\end{align}
	where $\alpha$ and $\lambda$ denote the path loss exponent and wavelength, respectively. $d$ is the distance between \textcolor{ black}{the} satellite and ground station, which can be obtained \textcolor{ black}{accordingly}:
	\begin{align}
	d = -r_{e}\sin(\varphi) + \sqrt{r_{e}^{2}\sin^2(\varphi) + h_{\mathrm{sat}}^{2} + 2r_{e}h_{\mathrm{sat}}},
	\label{eq:distance}
	\end{align}
	where $r_{e}$, $h_{\mathrm{sat}}$, and $\varphi$ \textcolor{ black}{represent} the radius of \textcolor{ black}{the} Earth, the altitude of \textcolor{ black}{the} satellite, and the elevation angle between the ground station and satellite, respectively. The path loss is proportional to $d^\alpha$; therefore, the elevation angle \textcolor{ black}{plays} an important role \textcolor{ black}{in} the path loss.
	
	\begin{figure}[t]
		\centering
		\includegraphics[width=\linewidth, page=2]{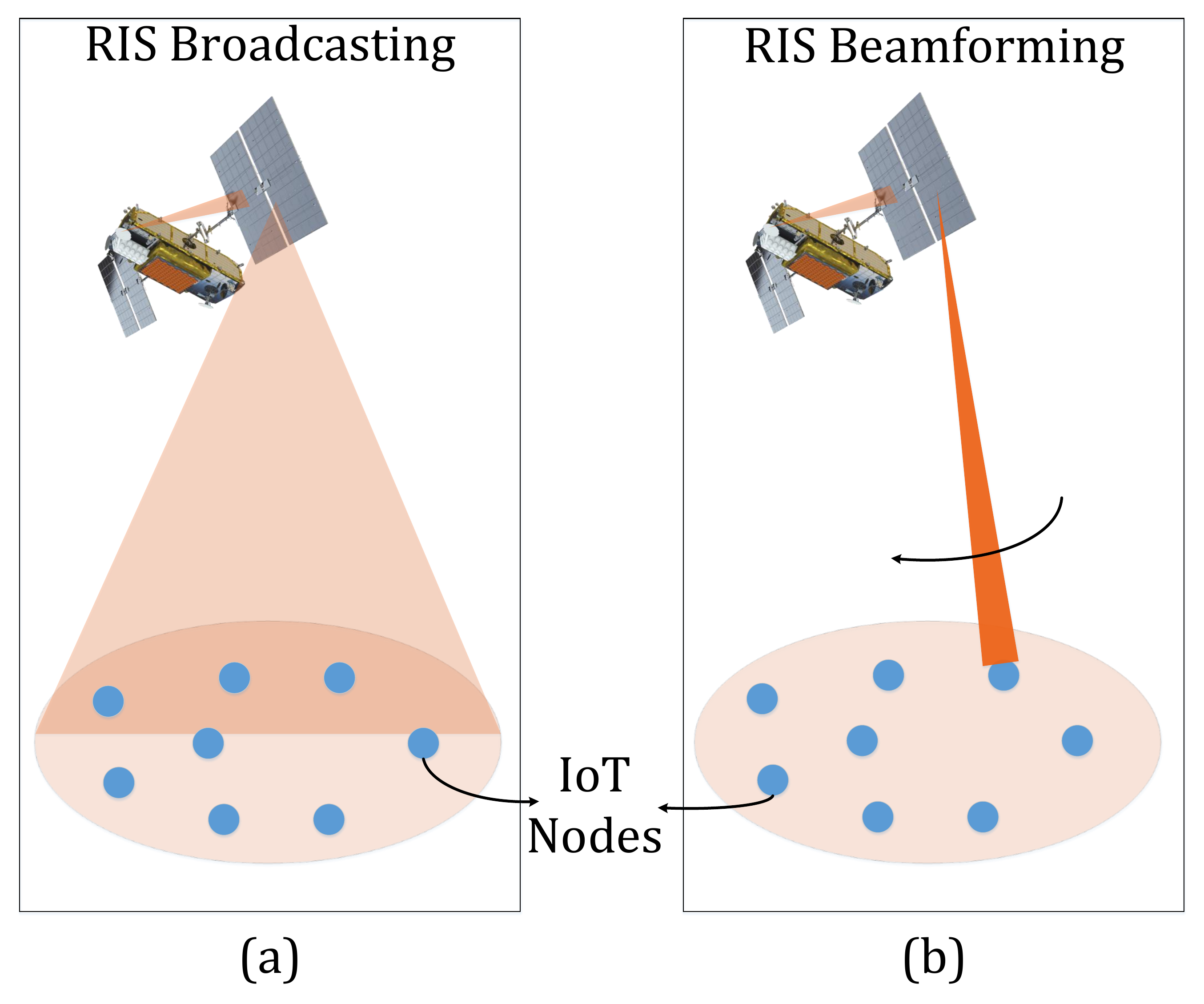}
		\caption{RIS-assisted satellites can serve in two modes: (a) RIS broadcasting and (b) RIS beamforming. \textcolor{ black}{In the former, the RISs scatter the incident wave; in the latter, the RISs perform specular reflection.}}
		\label{fig:boradcasting_beamforming}
	\end{figure}

	\subsection{Path Loss Models for \ac{RIS}-Assisted Communications}
	\textcolor{ black}{Here} we describe the path loss models for \ac{RIS}-assisted wireless communications in broadcasting and beamforming cases.
	
	\textcolor{ black}{We begin with} the broadcasting case as illustrated in \FGR{fig:boradcasting_beamforming}(a). If the transmitter is in the near-field of \ac{RIS}, and the surface is at least ten times larger electrically than the wavelength, $\lambda$, the surface scatters the incident spherical wave. Scattering \textcolor{ black}{the} incident wave creates a large beam, which can cover multiple ground stations at the same time. As the transmitter antenna is relatively near \textcolor{ black}{to} the \ac{RIS} deployed on the satellite, the scattering paradigm should be considered in the case of electrically large surfaces. The path loss model for \ac{RIS}-assisted wireless communications in the case of broadcasting is the following~\cite{tang2019wireless}:
	\begin{align}
	PL_{\text{BC}} \approx \frac{(4 \pi)^{\alpha}\left(d_{tx}+d_{rx}\right)^{\alpha}}{G_{t} G_{r} \lambda^{\alpha} A^{\alpha}},
	\end{align}
	where $d_{tx}$ and $d_{rx}$ are the distances from \textcolor{ black}{the} transmitter antenna to \textcolor{ black}{the} \ac{RIS} and from \textcolor{ black}{the} \ac{RIS} to \textcolor{ black}{the} receiver antenna, respectively. As $d_{rx}\gg d_{tx}$, the distance between \textcolor{ black}{the} transmitter antenna and \textcolor{ black}{the} \ac{RIS} can be omitted and $d_{rx} = d$. Then, the expression can be written as \textcolor{ black}{follows}:
	\begin{align}
	PL_{\text{BC}} \approx \frac{(4 \pi d)^{\alpha}}{G_{t} G_{r} \lambda^{\alpha} A^{\alpha}}.
	\label{eq:pl_bc}
	\end{align}
	\textcolor{ black}{Here,} $G_{t}$ and $G_{r}$ denote \textcolor{ black}{the} antenna gain for \textcolor{ black}{the} transmitter and receiver antennas, respectively. $A$ is the amplitude of the reflection coefficient of meta-atoms. It should be noted that the free-space path loss for \textcolor{ black}{the} \ac{RIS} broadcasting scheme is not dependent \textcolor{ black}{on} the number of \ac{RIS} elements, as seen in \eqref{eq:pl_bc}. In this scheme, \textcolor{ black}{the} \ac{RIS} scatters the incident wave. Thus, while a wider coverage is obtained, the amount of energy per area is slightly lower.
	
	\textcolor{ black}{But}, the transmitted wave can \textcolor{ black}{also} be focused on the receiver in order to improve the received signal quality\textcolor{ black}{,} as depicted in \FGR{fig:boradcasting_beamforming}(b). Both the transmitter and receiver can be in the far-field of the \ac{RIS}, or one of them can be in the near-field of the \ac{RIS}. The latter is the more appropriate model for satellite \ac{IoT} systems, which includes a transmitter \textcolor{ black}{near to the \ac{RIS}}. Therefore, we utilize the near-field beamforming scheme for \textcolor{ black}{the} \ac{RIS}. In other ways, it can be said that \ac{RIS} operates as a specular reflection~\cite{ellingson2019path}. The path loss model for near-field beamforming scheme in \textcolor{ black}{an} \ac{RIS}-assisted satellite is given as follows~\cite{tang2019wireless}:
	\begin{equation}
	PL_{\text{BF}}=\frac{64 \pi^{3}}{G_{t} G_{r} d_{x} d_{y} \lambda^{\alpha} A^{\alpha}\left|\sum_{n=1}^{N} \frac{\sqrt{F_{n}^{\text{combine }}}}{d_{tx} d_{rx}}\right|^{\alpha}},
	\label{eq:pl_bf}
	\end{equation}
	where $N$ is the number of meta-atoms. $d_{x}$ and $d_{y}$ are the size of each unit cell along the x-axis and y-axis, respectively. $F_{n}^{\text{combine}}$ denotes the accounted normalized power radiation pattern on the received signal power, \textcolor{ black}{which can be described accordingly:}
	\begin{equation}
	F_{n}^{\text {combine}}=F^{tx}\left(\theta_{n}^{tx}, \beta_{n}^{tx}\right) F\left(\theta_{n}^{t}, \beta_{n}^{t}\right) F\left(\theta_{n}^{r}, \beta_{n}^{r}\right) F^{rx}\left(\theta_{n}^{rx}, \beta_{n}^{rx}\right),
	\end{equation}
	where $F\left(\theta_{n}, \beta_{n}\right)$ shows the normalized radiation pattern for the elevation angle $\theta_{n}$ and the azimuth angle $\beta_{n}$ between \textcolor{ black}{the} \ac{RIS} and transmitter antenna (or receiver antenna). The normalized radiation pattern for the elevation angle $\theta$ and azimuth angle $\beta$ is defined as follows~\cite{balanis2016antenna}:
	\begin{equation}
	F(\theta, \beta)=\left\{\begin{aligned}
	\cos^{3}(\theta),\; \theta \in\left[0, \frac{\pi}{2}\right],\, \beta \in[0,2 \pi] \\
	0,\; \theta \in\left(\frac{\pi}{2}, \pi\right],\, \beta \in[0,2 \pi]
	\end{aligned}\right.
	\label{eq:f}
	\end{equation}
	It is worth noting that in order to maximize the received power, the transmitter antenna has to be deployed to \textcolor{ black}{the} satellite such that \textcolor{ black}{the transmitter antenna's} normal line is orthogonal to the surface. Namely, $\theta_{n}^{tx} = \theta_{n}^{t} = 0$ for all unit cells. Without loss of generality, it is assumed that $\theta_{n}^{rx} = \theta_{n}^{r} = \frac{\pi}{2} - \varphi$ and $d_{rx} \gg d_{tx}$. \textcolor{ black}{As expected, selecting $\theta_{n}^{tx}$ and $\theta_{n}^{t}$ is a design issue and it can be adjusted accordingly.} The most important issue in the \ac{RIS} beamforming scheme is that the loss gradually decreases with the increasing number of \textcolor{ black}{\ac{RIS}} elements.
	
	\subsection{Rain Attenuation}
	Rain attenuation is \textcolor{ black}{a} major propagation impairment for satellite systems. Rain can cause scattering and absorption of waves propagating through the atmosphere. Rain attenuation is described by ITU-R P.618-13 as~\cite{ITU-RP618-13}:
	\begin{equation}
	PL_{\text{rain}}=\xi_{R} L_{E} \;\;\text{(dB)},
	\end{equation}
	where $\xi_{R}$ and $L_{E}$ \textcolor{ black}{refer respectively to the specific frequency-dependent coefficient, described by ITU-R P.838~\cite{ITU-RP838}, and the effective path length.} First, we introduce the steps to find the value of $\xi_{R}$.
	\begin{equation}
	\xi_{R}=k\left(R_{0.01}\right)^{\nu} \quad (\mathrm{dB} / \mathrm{km}),
	\end{equation}
	where $R_{0.01}$ is the rainfall rate, \textcolor{ black}{which} can be obtained from ITU-R P.837~\cite{ITU-RP837} for the location of a ground station. $k$ and $\nu$ denote the frequency-dependent coefficients given in ITU-R P.838~\cite{ITU-RP838} as follows:
	\begin{equation}
	\begin{array}{l}
	k=\left[k_{H}+k_{V}+\left(k_{H}-k_{V}\right) \cos ^{2}(\varphi) \cos(2 \tau)\right] / 2 \\
	\nu=\left[k_{H} \nu_{H}+k_{V} \nu_{V}+\left(k_{H} \nu_{H}-k_{V} \nu_{V}\right) \cos ^{2}(\varphi) \cos(2 \tau)\right] / 2 \mathrm{k},
	\end{array}
	\end{equation}
	where $\tau = \frac{\pi}{4}$ for circular polarization\textcolor{ black}{,} and all coefficients are listed in \TAB{tab:parameters} for $4.25$ GHz and $10.5$ GHz.
	
	In order to find the effective path length $L_{E}$, which is
	\begin{equation}
	L_{\mathrm{E}}=L_{\mathrm{R}} v_{0.01} \quad(\mathrm{km}),
	\end{equation}
	where $v_{0.01}$ is the vertical adjustment factor modeled as follows:
	\begin{equation}
	v_{0.01}=\frac{1}{1+\sqrt{\sin (\varphi)}\left(31\left(1-\mathrm{e}^{-(\varphi /(1+\chi))}\right) \frac{\sqrt{L_{R} \xi_{R}}}{f^{2}}-0.45\right)}.
	\end{equation}
	$f$ is the frequency in GHz and $\chi$ is defined as~\cite{maral2020satellite}:
	\begin{equation}
	\chi=\left\{\begin{array}{ll}
	36 - |\text{latitude}|, \quad &|\text{latitude}|<36^{\circ} \\
	0, \quad &\text{otherwise.}
	\end{array}\right.
	\end{equation}
	Also, $L_{R}$ is described as:
	\begin{equation}
	L_{R} = \begin{cases} 
	\frac{L_{G} r_{0.01}}{\cos(\varphi)}, & \tan^{-1}\left(\frac{h_{R}-h_{S}}{L_{G}r_{0.01}}\right)>\varphi \\
	\frac{\left(h_{R}-h_{S}\right)}{\sin(\varphi)}, & \text{otherwise.}
	\end{cases}
	\end{equation}
	where $h_{S}$ is the altitude of the ground station in km. Also, $L_{G}$ is defined as:
	\begin{equation}
	L_{G} = \begin{cases} 
	L_{S}\cos(\varphi), & h_{R} - h_{S}>0 \\
	0, & h_{R} - h_{S} \leq 0
	\end{cases}
	\end{equation}
	where $L_{S}$ is a slant-path length calculated as follows:
	\begin{equation}
	L_{S} = \begin{cases} 
	\frac{2(h_{R} - h_{S})}{\sqrt{\sin^2(\varphi) + \frac{2(h_{R} - h_{S})}{r_e}} + \sin(\varphi)}, & \varphi\leq 5^{\circ} \\
	\frac{h_{R} - h_{S}}{\sin(\varphi)}, & \varphi > 5^{\circ}
	\end{cases}
	\end{equation}
	where $h_{R}$ is the effective height of the rain described by ITU-R~P.839~\cite{ITU-RP839} as:
	\begin{equation}
	h_{R} = h_{0} + 0.36 \quad (\textrm{km}),
	\end{equation}
	where $h_{0}$ is the mean $0^{\circ}$ isotherm height above mean sea level\textcolor{ black}{, which is a} site-specific value.
	
	\section{LEO Satellite-enabled IoT Networks}\label{sec:leo_system_model}
	
	\textcolor{ black}{\ac{LEO} satellite constellations have emerged as a popular way of assisting global \ac{IoT} networks with reasonable delays~\cite{wang2021joint}. Because of the expense of the gateway system and inflexible infrastructure, direct access is desired in LEO satellite IoT systems~\cite{fraire2019direct}.} In this section, we describe a novel energy-efficient system model for satellite communications with a transmitter antenna\footnote{It should be noted that it is a receiver antenna for uplink communications.} in the near-field of the \ac{RIS} on a satellite. In order to maximize the received power, the transmitter antenna \textcolor{ black}{needs} to be aligned with the normal line of the \ac{RIS}, which makes the angle between the normal line and antenna beam zero. Thus, the normalized radiation pattern takes \textcolor{ black}{the} maximum value as observed in \eqref{eq:f}. Furthermore, the distance $d_{tx}$ between \textcolor{ black}{the} transmit antenna and the \ac{RIS} is very short compared to the distance to ground station. Hence, the wireless channel between \textcolor{ black}{the} transmitter and \ac{RIS} can be omitted. The received signal $y$ can be introduced as \textcolor{ black}{follows:}
	\begin{equation}
	y =\sqrt{\frac{P_{t}}{PL}} \mathbf{g^\mathrm{T}} \mathbf{\Phi} \mathbf{h} x + w,
	\end{equation}
	where $x$ and $w$ stand for the transmitted signal with power $P_{t}$ and \ac{AWGN} at \textcolor{ black}{the} receiver, respectively. The noise term, $w$, can be assumed to be distributed with $\mathcal{C} \mathcal{N}\left(0, N_{0}\right)$. $\mathbf{h}$ is the channel coefficient vector for the link between \textcolor{ black}{the} \ac{RIS} and receiver, such that $\mathbf{h}=\left[h_{1}, h_{2}, \ldots, h_{N}\right]$. $\mathbf{g}$ stands for the channel coefficient for the link between \textcolor{ black}{the} transmitter and \ac{RIS}. As \textcolor{ black}{the} transmit antenna is close to \textcolor{ black}{the} \ac{RIS}, we can ignore the channel effects in between \textcolor{ black}{the} transmitter and \ac{RIS}. Therefore, $\mathbf{g}$ can be selected as the vector of all ones such that $\mathbf{g} = \mathbf{1}_{N}$. \textcolor{ black}{It should be noted that satellite communications generally have \ac{LOS} propagations, and thus, a Rician fading model is employed in satellite channel models~\cite{letzepis2008capacity}. We therefore assume that the channel coefficients follow a Rician distribution. As in~\cite{you2020massive}, the Rician factor of the channel is selected as $K = 10$.} $\mathbf{\Phi}$ denotes the \textcolor{ black}{responses of the RIS element can be expressed as follows:}
	\begin{equation}
	\Phi=\operatorname{diag}\left\{A_{1} \mathrm{e}^{j \phi_{1}}, \ldots, A_{N} \mathrm{e}^{j \phi_{N}}\right\},
	\end{equation}
	where $A_{i}$ and $\phi_{i}$ are the amplitude and phase response of $i$-th \ac{RIS} element, \textcolor{ black}{respectively}. Throughout this study, \acp{RIS} are assumed \textcolor{ black}{to be lossless}. Therefore, $A_{i} = A = 1, \; \forall i$. 
	
	Next, regarding the received signal, the instantaneous \ac{SNR} $\gamma$ can be given as \textcolor{ black}{follows}:
	\begin{equation}
	\gamma = \frac{\left|\mathbf{g^\mathrm{T}}\Phi \mathbf{h}\right|^{2}P_{t}}{N_{0}PL},
	\label{eq:snr}
	\end{equation}
	where $PL$ \textcolor{ black}{represents} the total loss, including free-space path loss\footnote{It can be either RIS-assisted wireless communications or not. For RIS-assisted communications, $PL_{\text{BC}}$ and $PL_{\text{BF}}$ are employed for RIS broadcasting and RIS beamforming, respectively. \textcolor{ black}{By contrast}, it is equal to $PL_{\text{FS}}$ in case of wireless communications without \ac{RIS}.} and rain attenuation. \textcolor{ black}{As we can see in (\ref{eq:snr}), the increased number of RIS elements improves the received SNR value. However, it should be emphasized that the increase in SNR values with respect to the number of RIS elements heavily depends on RIS design, as detailed above. Therefore, this work employs an RIS prototype, and its model is given in~\cite{dai2020reconfigurable}, rather than using a hypothetical RIS model.} 
	
	For \textcolor{ black}{the} free-space path loss calculation, \eqref{eq:pl} is utilized for satellite communications \textcolor{ black}{in non-}\ac{RIS} \textcolor{ black}{case}. In \ac{RIS}-assisted case, there two different modes, which are related to the dimension of the \ac{RIS} element unit. For \textcolor{ black}{the} \ac{RIS}-assisted broadcasting scheme, \eqref{eq:pl_bc} gives the free-space path loss, while \eqref{eq:pl_bf} is the free-space path loss expression for \ac{RIS}-assisted beamforming. Considering the \ac{SNR}, the achievable data rate can be expressed as \textcolor{ black}{follows:}
	\begin{equation}
	R =\log _{2}\left(1+\gamma\right)\quad \text{(bits/s/Hz)}.
	\end{equation}
	\textcolor{ black}{Due to the rise in the SNR value with an increasing number of RIS elements, the achievable rate also improves.}
	
	\section{Numerical Results and Discussion}\label{sec:results}
	
	
	\begin{table}[!t]
		\centering
		\caption{Frequency-dependent coefficients which used in rain attenuation calculations for circular polarization \\(ITU-R Rec. P.838).}
		\begin{tabular}{lcc}
			\toprule \toprule
			\multicolumn{1}{l}{\textbf{Coefficients}} & \multicolumn{1}{c}{\textbf{4.25 GHz}} & \multicolumn{1}{c}{\textbf{10.5 GHz}} \\  \midrule
			\multicolumn{1}{l}{$k_H$}       & $7.3420\mathrm{e}\minus4$      & $1.1926\mathrm{e}\minus2$   \\
			\multicolumn{1}{l}{$k_V$}       & $6.8259\mathrm{e}\minus4$      & $1.0526\mathrm{e}\minus2$   \\
			\multicolumn{1}{l}{$\nu_H$}  & $1.1489$                  & $1.2602$               \\
			\multicolumn{1}{l}{$\nu_V$}  & $1.1034$                  & $1.2469$               \\ 
			\multicolumn{1}{l}{Polarization}  & Circular                  & Circular               \\ \midrule
			\multicolumn{1}{l}{$k$}         & $7.084\mathrm{e}\minus4$       & $1.123\mathrm{e}\minus2$    \\
			\multicolumn{1}{l}{$\nu$}    & $1.127$                   & $1.254$                \\ \bottomrule \bottomrule
		\end{tabular}
		\label{tab:parameters}
	\end{table}
	
	\begin{table}[!t]
		\centering
		\caption{Location-specific coefficients which used in rain attenuation calculations for circular polarization \\(ITU-R Rec. P.838 \& ITU-R Rec. P.839).}
		\begin{tabular}{lcc}
			\toprule \toprule
			\multicolumn{1}{l}{\textbf{Parameters}} & \multicolumn{1}{c}{\textbf{Values}}  \\  \midrule
			\multicolumn{1}{l}{Location}       & Istanbul, Turkey   \\
			\multicolumn{1}{l}{Latitude}       & $41^{\circ}$N      \\
			\multicolumn{1}{l}{Longitude}      & $29^{\circ}$E      \\
			\multicolumn{1}{l}{$h_{0}$}        & $2.53$ km            \\ 
			\multicolumn{1}{l}{$R_{0.01}$}     & $31.119$           \\
			\multicolumn{1}{l}{$h_{S}$}        & $1$ m              \\ 
			\multicolumn{1}{l}{$h_{sat}$}      & $800$ km              \\
			\multicolumn{1}{l}{$r_{e}$}        & $6371$ km              \\ \bottomrule \bottomrule
		\end{tabular}
		\label{tab:loc_parameters}
	\end{table}

	In this section, we \textcolor{ black}{present a} comprehensive simulation and discuss simulation results. First, we focus on \textcolor{ black}{the} downlink capacity for satellite \ac{IoT} systems. Then, we investigate \textcolor{ black}{the} uplink capacity of satellite-supported \ac{IoT} communications. 
	
	\subsection{\textcolor{ black}{Simulation Parameters}}
	We consider the simulation results for the two prominent bands for satellite \ac{IoT} systems, namely C- and X-band. In simulations, the path loss model given in \eqref{eq:pl} \textcolor{ black}{was} used for the case without \acp{RIS}, while the models given \textcolor{ black}{in \eqref{eq:pl_bc} and \eqref{eq:pl_bf} are for RIS broadcasting and RIS beamforming,} respectively.
	
	As the rain attenuation changes \textcolor{ black}{significantly} with respect to the operating frequency, it is important to properly acquire the frequency-dependent coefficients such as $k$ and $\nu$. For example, $k$ is found for C- and X-band as $7.084\mathrm{e}\minus4$ and $1.123\mathrm{e}\minus2$, respectively. Besides, $\nu$ is equal to $1.127$ and $1.254$ for C- and X-band, respectively. \textcolor{ black}{It is important to note} that these values are valid for circular polarization, i.e. $\tau = \frac{\pi}{4}$. All frequency-dependent coefficients are given in \TAB{tab:parameters}.
	
	Also, in \textcolor{ black}{the} simulations, we assume that the ground stations (i.e. IoT devices) are located in Istanbul, Turkey, which is located at $41^{\circ}$ north latitude and $29^{\circ}$ east longitude. Site-specific coefficients such as the mean $0^{\circ}$ isotherm height above mean sea level $h_{0}$ and the rainfall rate $R_{0.01}$ is found as $2.53$ km and $31.119$ for Istanbul, Turkey. These parameters are summarized in \TAB{tab:loc_parameters}. As we investigate the \ac{LEO} satellites, the satellite altitude $h_{sat}$ is chosen as $800$ km.
	
	\textcolor{ black}{Other crucial factors to consider are the characteristics of the transmitter and receiver antennas and the RIS units. In the simulations, we employed} the antennas and \acp{RIS} given in~\cite{tang2019wireless}. These antennas have the normalized radiation pattern $F(\theta, \beta)$ \textcolor{ black}{that can be defined as follows}\footnote{Here, the normalized radiation pattern is given only for \textcolor{ black}{the} transmit antenna. It should be noted that this expression can also be utilized for \textcolor{ black}{the} receive antenna.}: 
	\begin{equation}
	F^{tx}(\theta, \beta) = \begin{cases} 
	\cos^{13}(\theta) , & f = 4.25\; \textrm{GHz}\; \textrm{(C-band)} \\
	\cos^{62}(\theta), & f = 10.5\; \textrm{GHz}\; \textrm{(X-band)}.
	\end{cases}
	\end{equation}
	Furthermore, antenna gains are given as $14.5$ dB and $21$ dB for C-band and X-band antennas, respectively. As \textcolor{ black}{discussed} above, the distance between \textcolor{ black}{the} transmit antenna and the surface is short. Hence, $d_{tx}$ is selected as $1$ m to keep \textcolor{ black}{the} near-field condition for both bands. $d_{rx}$ is equal to the distance given by \eqref{eq:distance}. In addition, \ac{RIS} elements are square and their edge lengths are $0.012$ m and $0.01$ m for $4.25$ GHz and $10.5$ GHz, respectively.
	
	The path loss exponent $\alpha$ is chosen as $2$, which is a generally accepted value. Additionally, the small-scale fading for the channel between \textcolor{ black}{the} \ac{RIS} and receiver is modeled by Rice distribution with the shape parameter of $K = 10$ in order to allow \textcolor{ black}{a} few \ac{NLOS} paths. Last, the effective noise power for the overall system is chosen as $\minus 100$ dB. \TAB{tab:path_loss_parameters} summarizes the parameters employed in \textcolor{ black}{calculating the} free-space path loss.

	\begin{table}[!t]
		\centering
		\caption{Parameters for free-space path loss calculation for \acp{RIS} and antennas operating at C-band and X-band.}
		\begin{tabular}{lcc}
			\toprule \toprule
			\multicolumn{1}{l}{\textbf{Parameters}} & \multicolumn{1}{c}{\textbf{4.25 GHz}} & \multicolumn{1}{c}{\textbf{10.5 GHz}} \\  \midrule
			\multicolumn{1}{l}{$d_{tx}$}       & $1$ m     & $1$ m   \\
			\multicolumn{1}{l}{$d_{rx}$}       & $d$      & $d$   \\
			\multicolumn{1}{l}{$\alpha$}     & $2$                       & $2$               \\
			\multicolumn{1}{l}{$d_{x}$~\cite{tang2019wireless}}     & $0.012$ m                  & $0.01$ m                \\ 
			\multicolumn{1}{l}{$d_{y}$~\cite{tang2019wireless}}     & $0.012$ m                  & $0.01$ m                \\ 
			\multicolumn{1}{l}{$G_{t}$~\cite{tang2019wireless}}     & $14.5$ dB      & $21$ dB   \\
			\multicolumn{1}{l}{$G_{r}$~\cite{tang2019wireless}}     & $14.5$ dB      & $21$ dB   \\
			\multicolumn{1}{l}{$F^{tx}(\theta, \beta)$~\cite{tang2019wireless}}         & $\cos^{13}(\theta)$       & $\cos^{62}(\theta)$    \\
			\multicolumn{1}{l}{$F^{rx}(\theta, \beta)$~\cite{tang2019wireless}}         & $\cos^{13}(\theta)$       & $\cos^{62}(\theta)$    \\
			\multicolumn{1}{l}{$\theta^{tx}$}         & $0$       & $0$    \\
			\multicolumn{1}{l}{$\theta^{t}$}          & $0$       & $0$    \\
			\multicolumn{1}{l}{$\theta^{rx}$}         & $\frac{\pi}{2}-\varphi$       & $\frac{\pi}{2}-\varphi$    \\
			\multicolumn{1}{l}{$\theta^{r}$}         & $\frac{\pi}{2}-\varphi$       & $\frac{\pi}{2}-\varphi$    \\
			\multicolumn{1}{l}{$K$}         & $10$       & $10$    \\
			\multicolumn{1}{l}{$N_{0}$}         & $\minus100$ dB      & $\minus100$ dB    \\ \bottomrule \bottomrule
		\end{tabular}
		\label{tab:path_loss_parameters}
	\end{table}

	\subsection{Downlink Performance Analysis}
	
	\begin{figure*}[!t]
		\centering
		\subfigure[]{
			\label{fig:capacity_vs_pt_with_without_ris_c_band_dl}
			\includegraphics[width=0.3\linewidth]{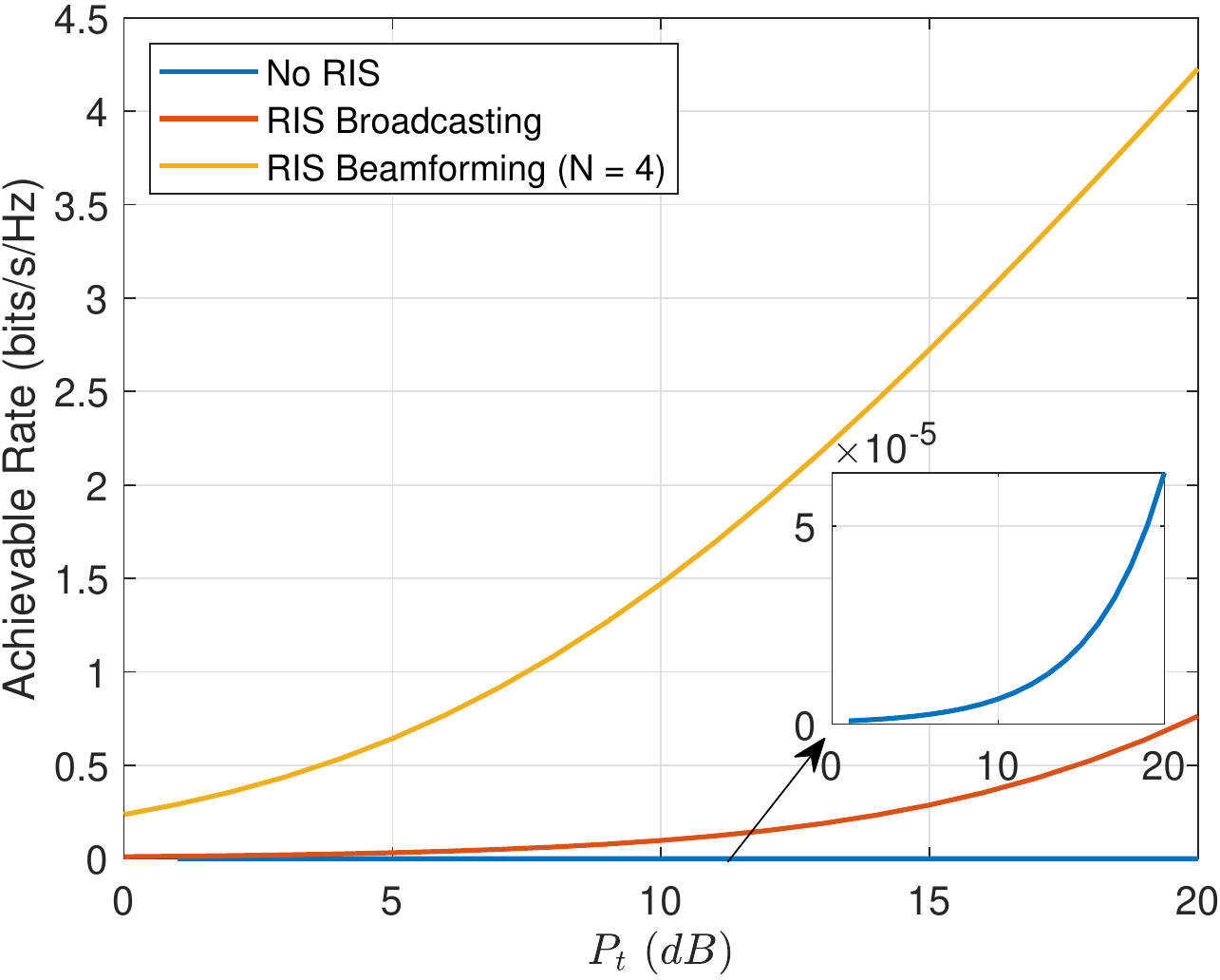}}
		\subfigure[]{
			\label{fig:capacity_vs_pt_c_band_dl}
			\includegraphics[width=0.3\linewidth]{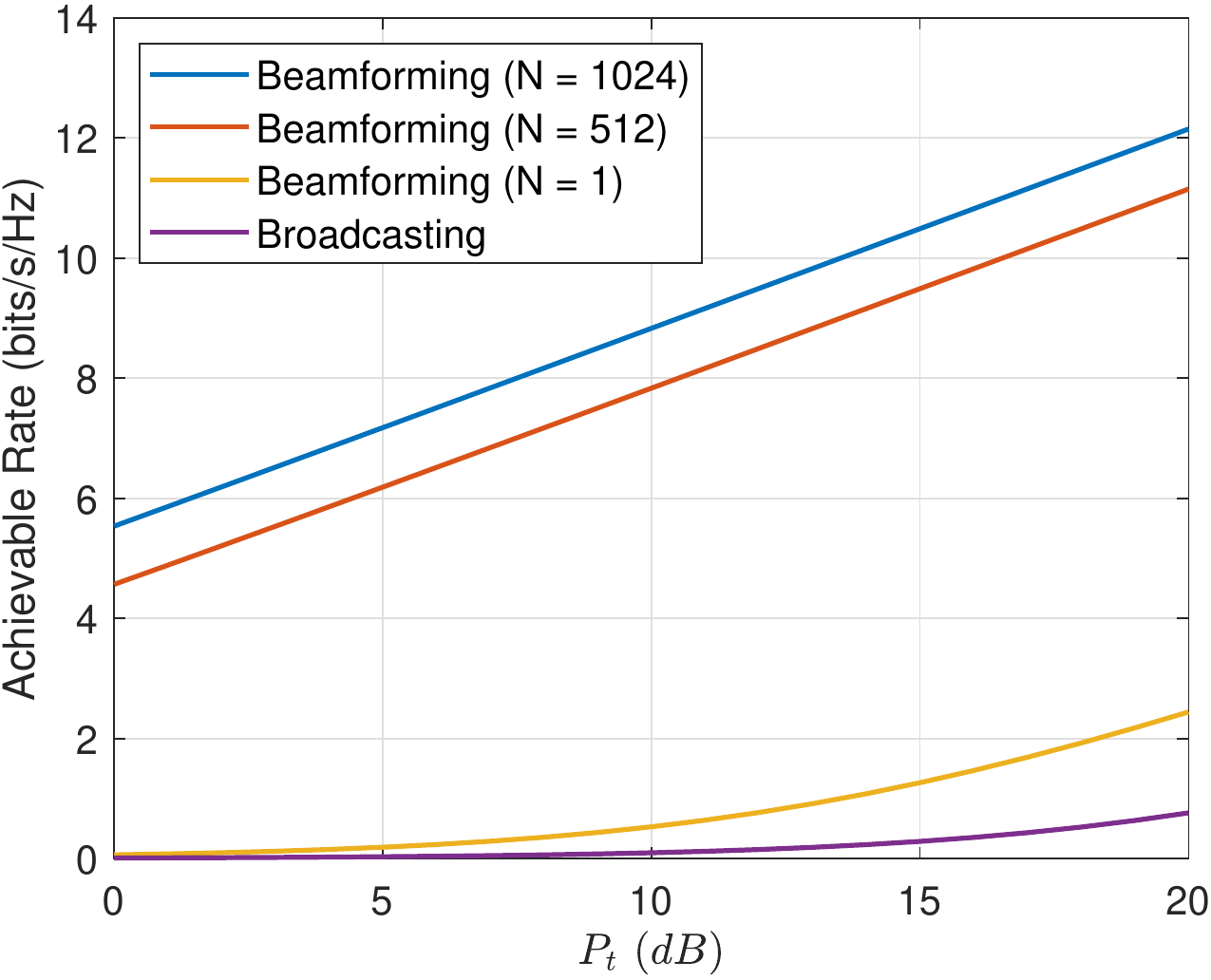}}
		\subfigure[]{
			\label{fig:capacity_vs_pt_x_band_dl}
			\includegraphics[width=0.3\linewidth]{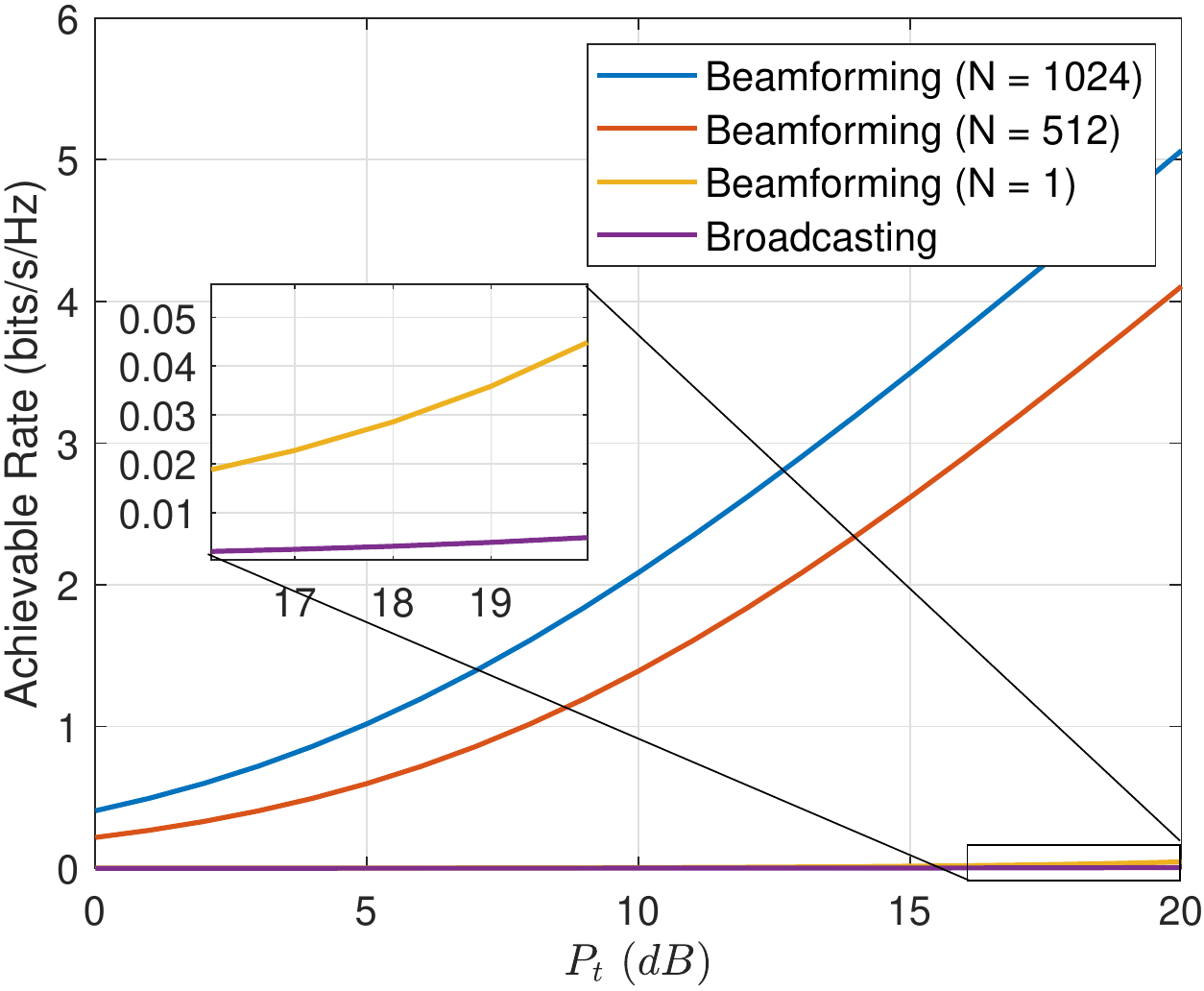}}
		\caption{Achievable downlink rate \textcolor{ black}{relative to} transmit power $P_t$ for (a) \textcolor{ black}{the} comparison between \textcolor{ black}{the case without RISs} and RIS-assisted satellites, (b) the various number of RIS elements in C-band, and (c) the various number of RIS elements in X-band. It should be noted that the elevation angle is $\frac{\pi}{2}$.}
		\label{fig:dl_results_pt}
	\end{figure*}

	\begin{figure*}[!t]
		\centering
		\subfigure[]{
			\label{fig:capacity_vs_elevation_with_without_ris_c_band_dl}
			\includegraphics[width=0.3\linewidth]{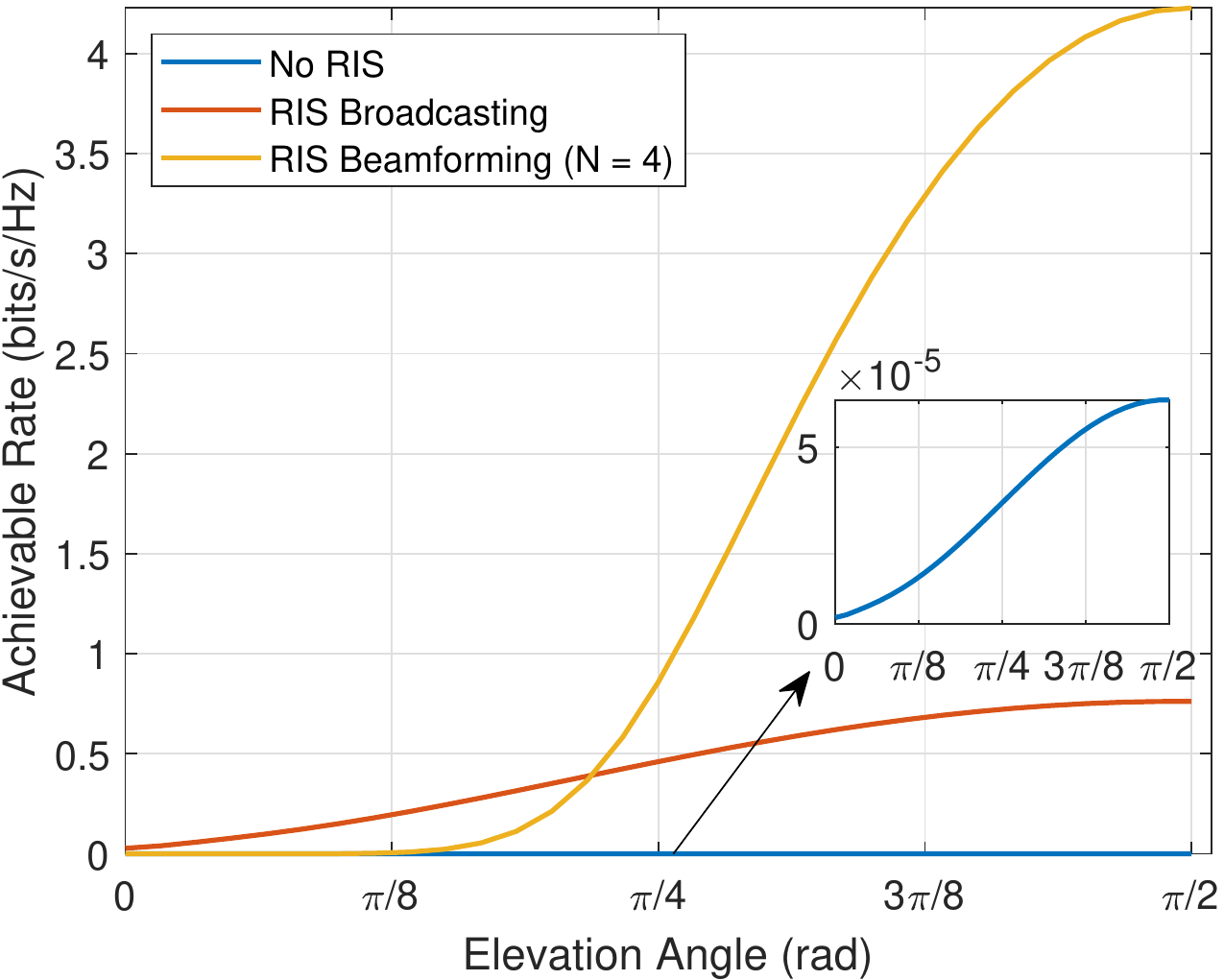}}
		\subfigure[]{
			\label{fig:capacity_vs_elevation_c_band_dl}
			\includegraphics[width=0.3\linewidth]{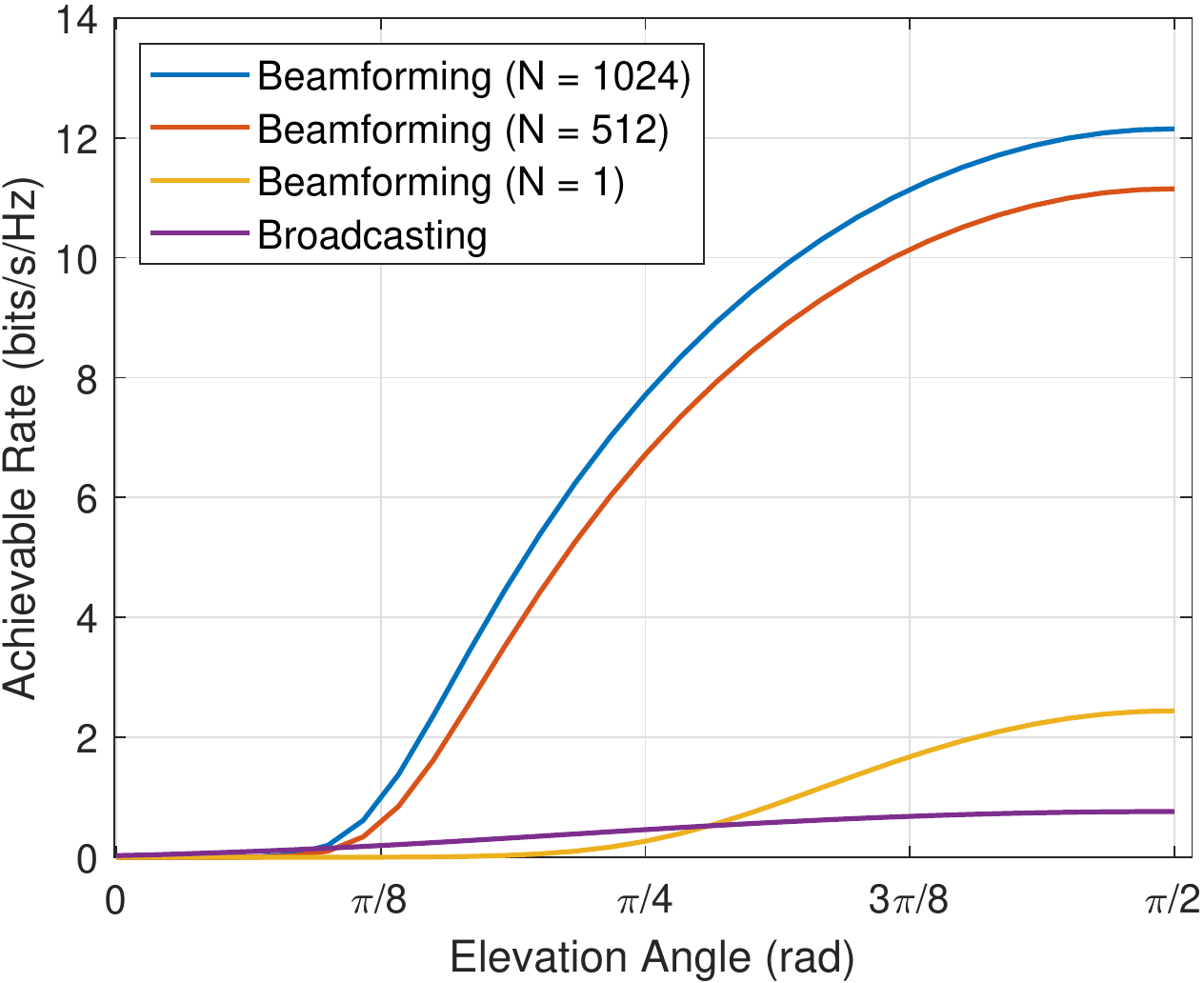}}
		\subfigure[]{
			\label{fig:capacity_vs_elevation_x_band_dl}
			\includegraphics[width=0.3\linewidth]{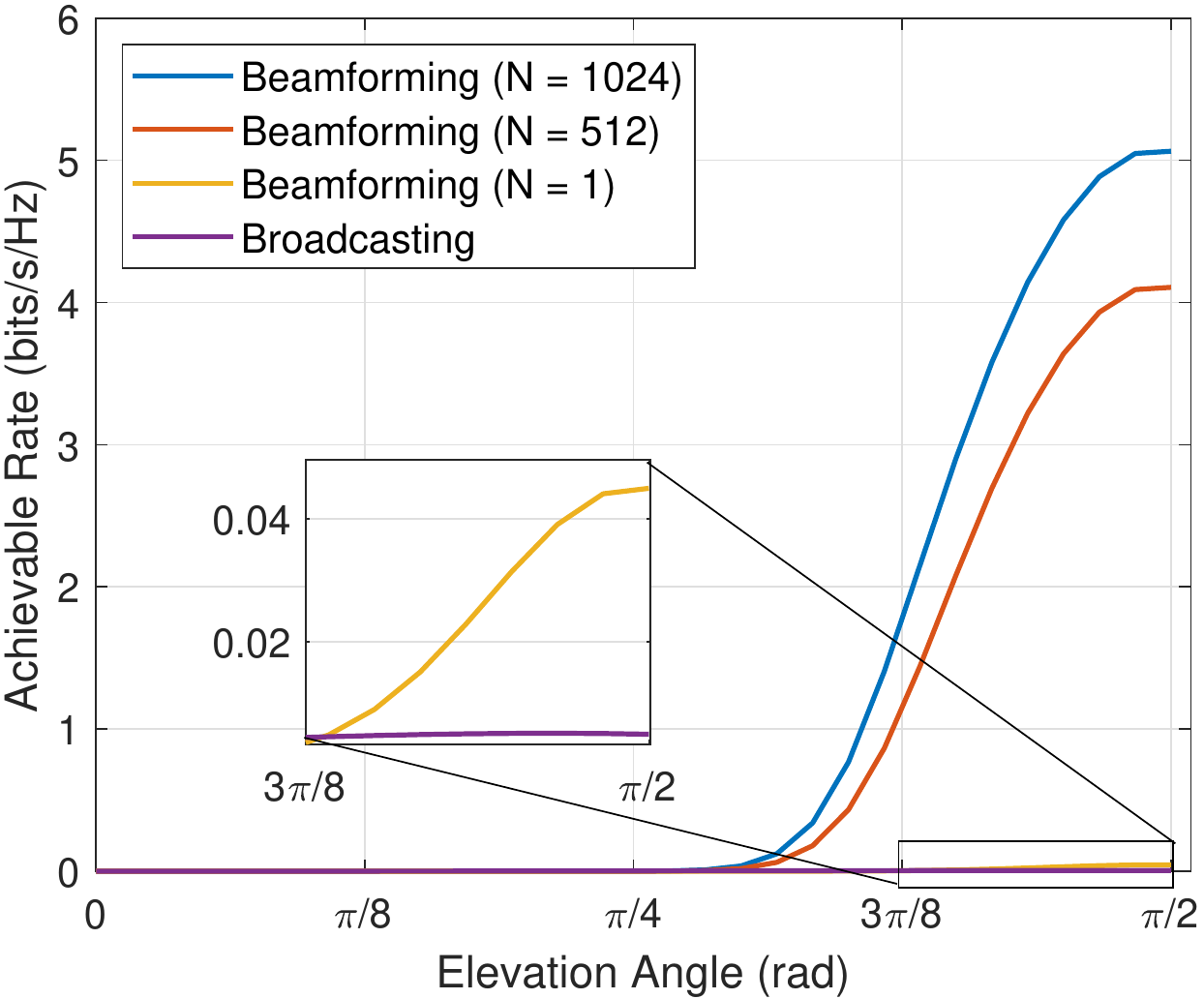}}
		\caption{Achievable downlink rate \textcolor{ black}{relative to} elevation angle $\varphi$ for (a) \textcolor{ black}{the} comparison between \textcolor{ black}{the case without RISs} and RIS-assisted satellites, (b) the various number of RIS elements in C-band, and (c) the various number of RIS elements in X-band. It should be noted that the transmit power is $100$ W.}
		\label{fig:dl_results_elevation}
	\end{figure*}

	\textcolor{ black}{Here} we investigate the downlink performance of satellite \ac{IoT} networks in terms of achievable data rate in three cases: \textcolor{ black}{(i)} without \ac{RIS}, \textcolor{ black}{(ii)} \ac{RIS} broadcasting, and \textcolor{ black}{(iii)} \ac{RIS} beamforming. First, we compare the achievable data rate for conventional satellite systems with \ac{RIS}-assisted satellites in~\FGR{fig:capacity_vs_pt_with_without_ris_c_band_dl} for C-band. The simulation results show that \ac{RIS}-assisted satellites provide much higher capacity than conventional satellites, regardless of which scheme they \textcolor{ black}{use}. In fact, the \ac{RIS} broadcasting scheme \textcolor{ black}{is shown to provide an achievable rate of up to $10^4$ times higher than the case without RISs. RIS beamforming is shown to provide a rate of up to $10^5$ times higher.} In other words, the higher capacity can be \textcolor{ black}{achieved} with a lower transmit power by using \ac{RIS} \textcolor{ black}{units on} satellites. Next, we investigate the impact of the number of \ac{RIS} elements in~\FGR{fig:capacity_vs_pt_c_band_dl}. 
	
	As the number of \ac{RIS} elements increases, the achievable capacity for the \ac{RIS} beamforming case increases, as expected. However, as given in \eqref{eq:pl_bc}, in the case of \ac{RIS} broadcasting, the number of elements does not affect the performance. Due to specular reflection, \ac{RIS} beamforming even with a single element can achieve \textcolor{ black}{a higher} data rate than the broadcasting case. The main reason \textcolor{ black}{for this is} the scattering of energy \textcolor{ black}{over} a wide area. The broadcasting scheme can support more \ac{IoT} devices than the beamforming scheme because of \textcolor{ black}{the} larger coverage area provided by the broadcasting. Last, as rain attenuation coefficients and the normalized radiation pattern \textcolor{ black}{depend heavily} on the operating frequency. The simulation results are \textcolor{ black}{shown} in \FGR{fig:capacity_vs_pt_x_band_dl}. For large $N$ values, the achievable rate decreases to less than half, while the ratio of performance loss increases as the number of elements decreases. For example, in the case of a single element, the achievable rate decreases to almost one-fourth of that in the C-band.
	
	\textcolor{ black}{As LEO satellites orbit the Earth,} the distance between them and the ground stations varies depending on the elevation angle. Distance is one of the major contributors to free-space path loss. Therefore, we have performed the simulations by considering varying elevation angles between zero and $\frac{\pi}{2}$ rad. \FGR{fig:dl_results_elevation} \textcolor{ black}{shows} that the RIS beamforming \textcolor{ black}{performs better} than broadcasting and \textcolor{ black}{the case without \acp{RIS}}. Since $\theta^{rx}$ and $\theta^{r}$ are defined as $\theta^{rx} = \theta^{r} = \frac{\pi}{2} - \varphi$, the elevation angle changes the normalized received radiation pattern. Therefore, \textcolor{ black}{with fewer} RIS elements at lower elevation angles, the RIS broadcasting scheme provides \textcolor{ black}{a} slightly better achievable rate than RIS broadcasting. However, increasing the number of RIS elements compensates for the path loss increase owing to the increased distance and decreased received radiation pattern as seen in~\FGR{fig:capacity_vs_elevation_c_band_dl} and~\FGR{fig:capacity_vs_elevation_x_band_dl}. 
	
	\vspace{-0.2cm}

	\subsection{Uplink Performance Analysis}
	
	\begin{figure*}[!t]
		\centering
		\subfigure[]{
			\label{fig:capacity_vs_pt_with_without_ris_c_band_ul}
			\includegraphics[width=0.31\linewidth]{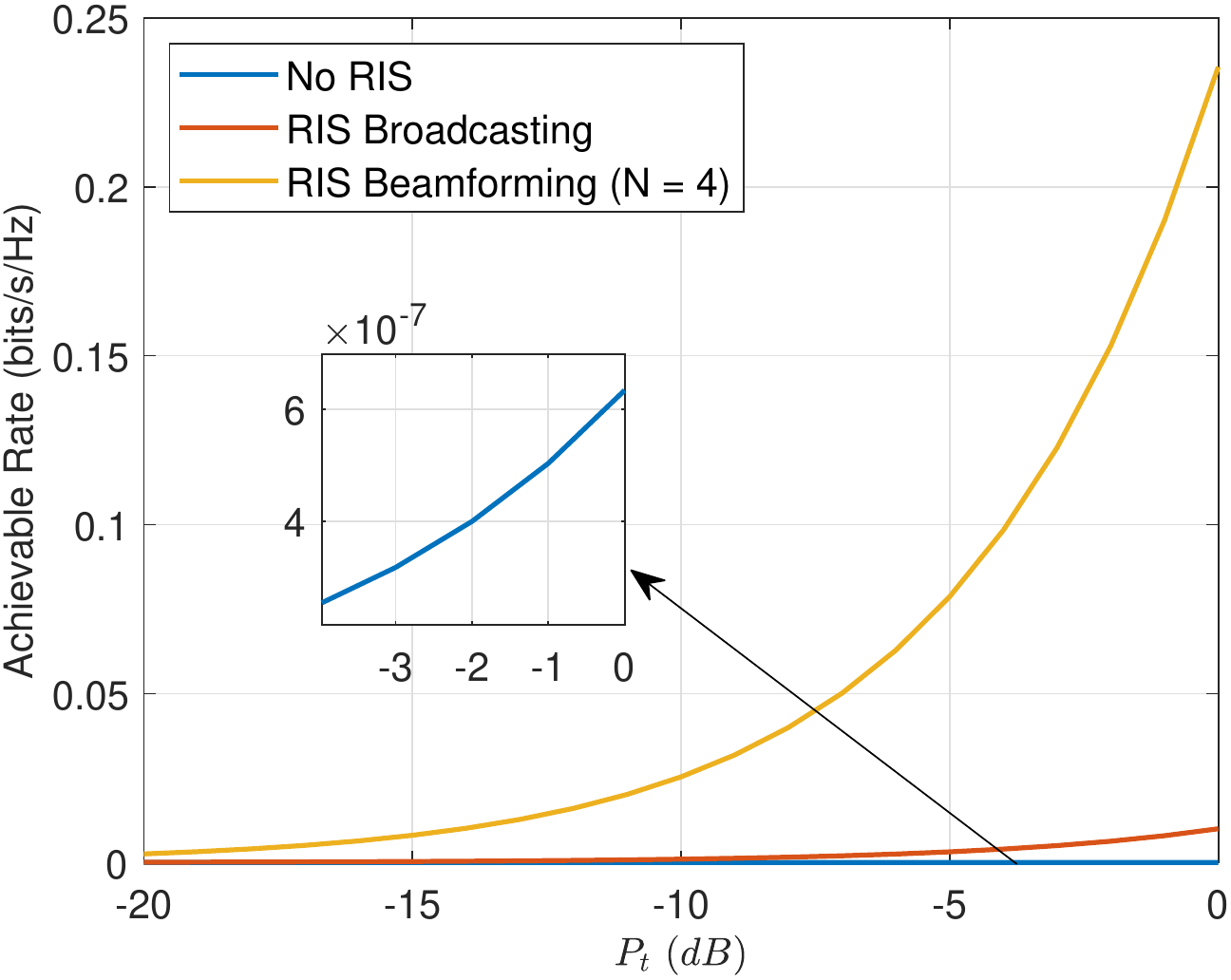}}
		\subfigure[]{
			\label{fig:capacity_vs_pt_c_band_ul}
			\includegraphics[width=0.298\linewidth]{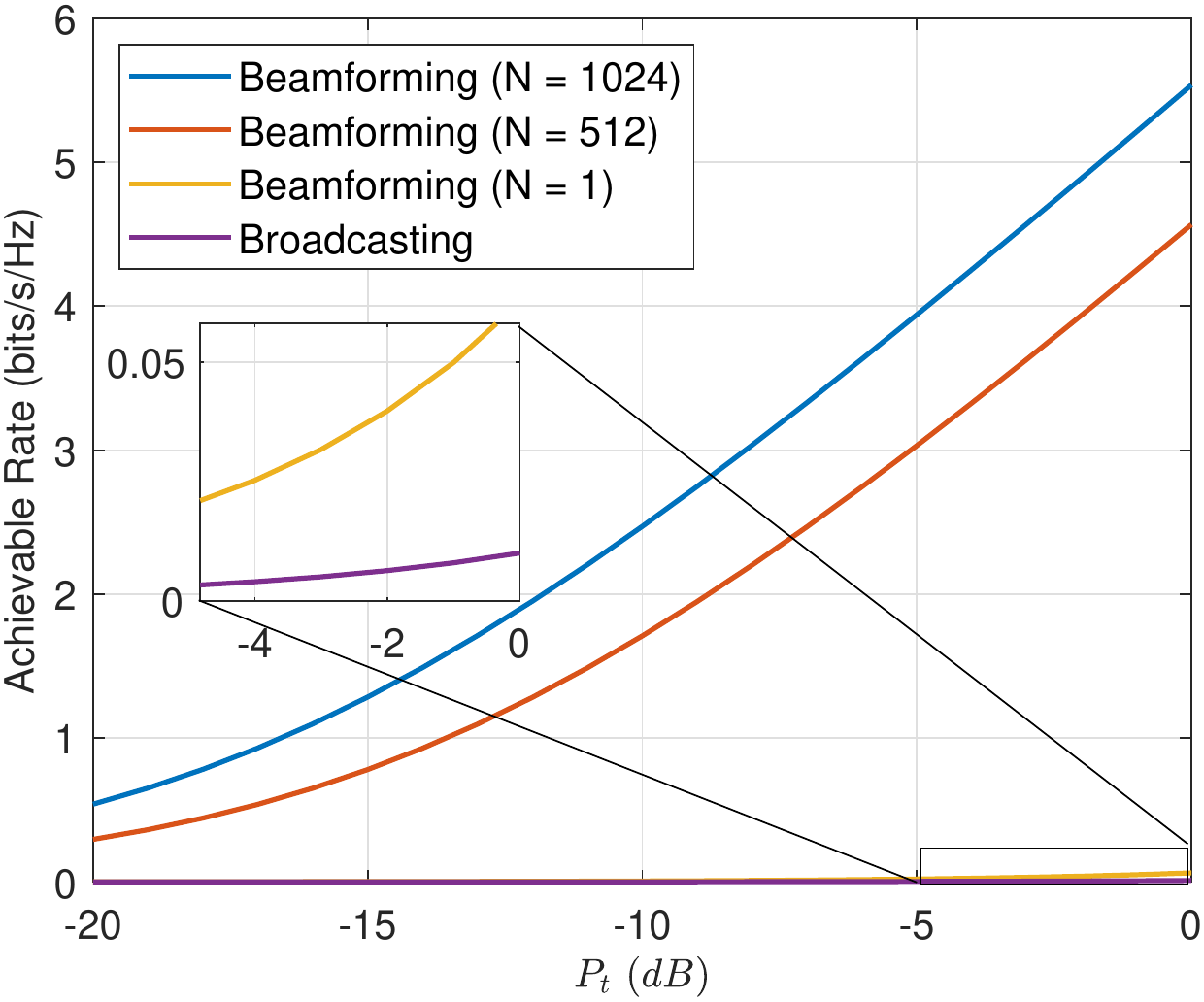}}
		\subfigure[]{
			\label{fig:capacity_vs_pt_x_band_ul}
			\includegraphics[width=0.31\linewidth]{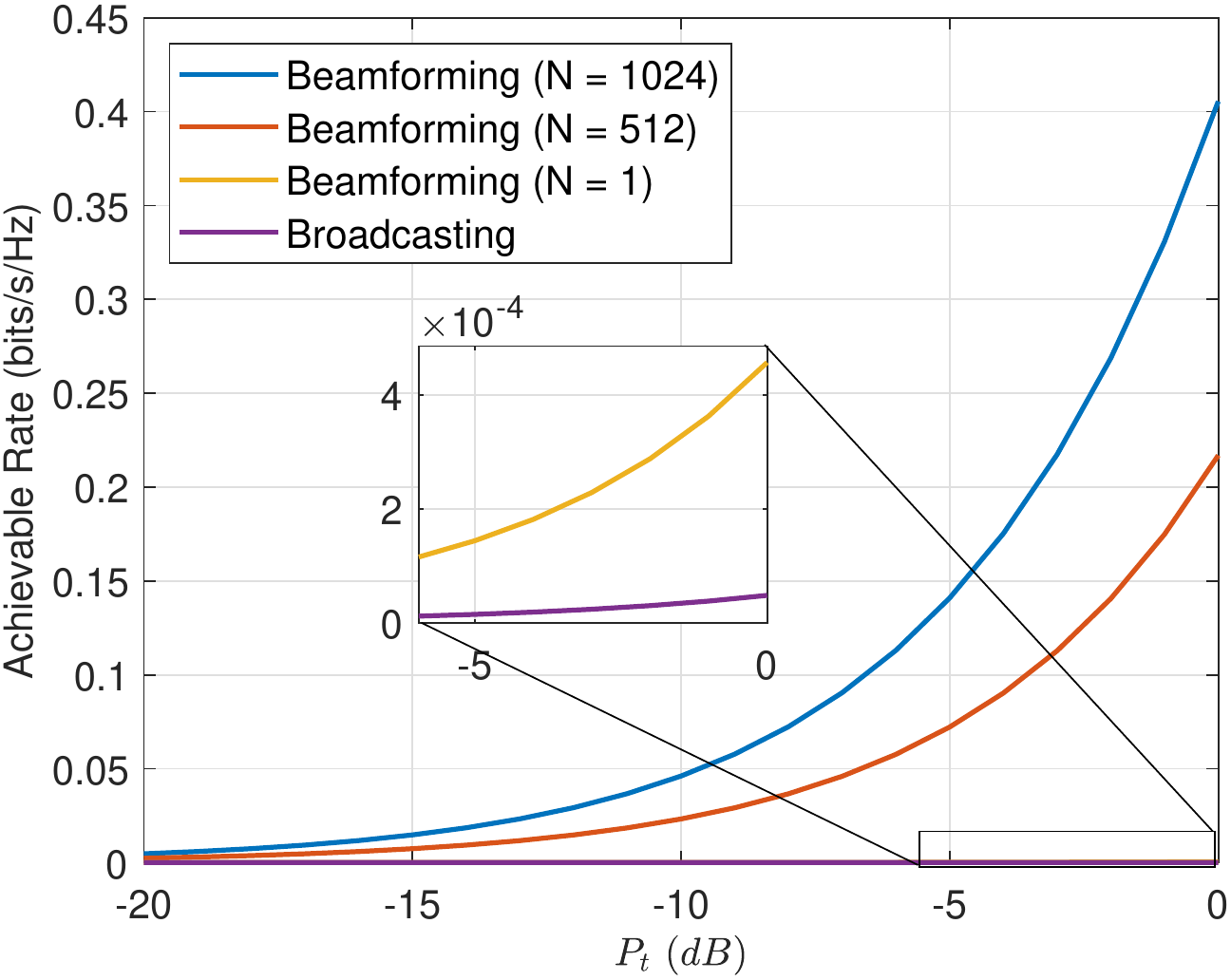}}
		\caption{Achievable uplink rate \textcolor{ black}{relative to} transmit power $P_t$ for (a) \textcolor{ black}{the} comparison between \textcolor{ black}{the case without RISs} and RIS-assisted satellites, (b) the various number of RIS elements in C-band, and (c) the various number of RIS elements in X-band. It should be noted that the elevation angle is $\frac{\pi}{2}$.}
		\label{fig:ul_results_pt}
	\end{figure*}

	\begin{figure*}[!t]
		\centering
		\subfigure[]{
			\label{fig:capacity_vs_elevation_with_without_ris_c_band_ul}
			\includegraphics[width=0.31\linewidth]{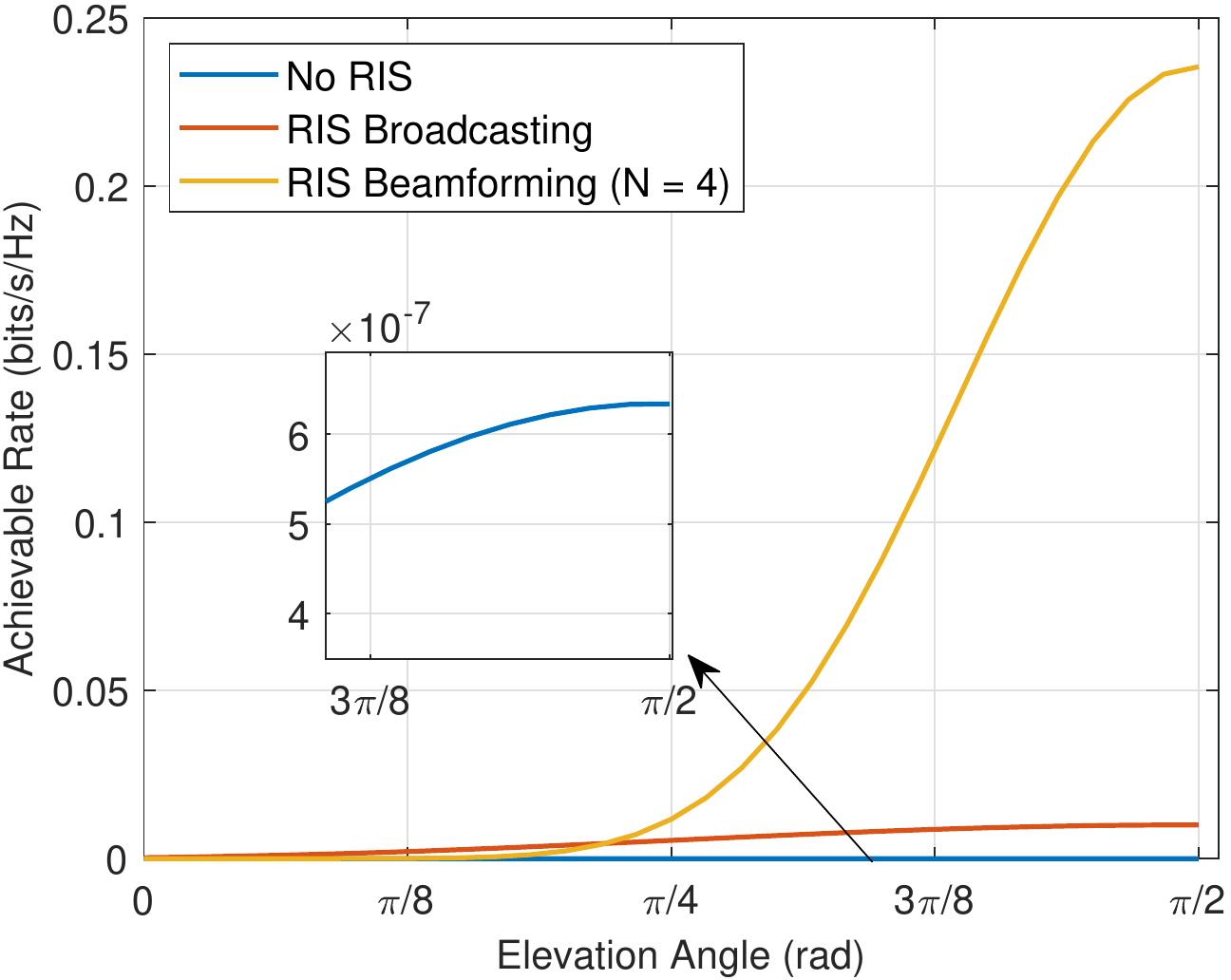}}
		\subfigure[]{
			\label{fig:capacity_vs_elevation_c_band_ul}
			\includegraphics[width=0.298\linewidth]{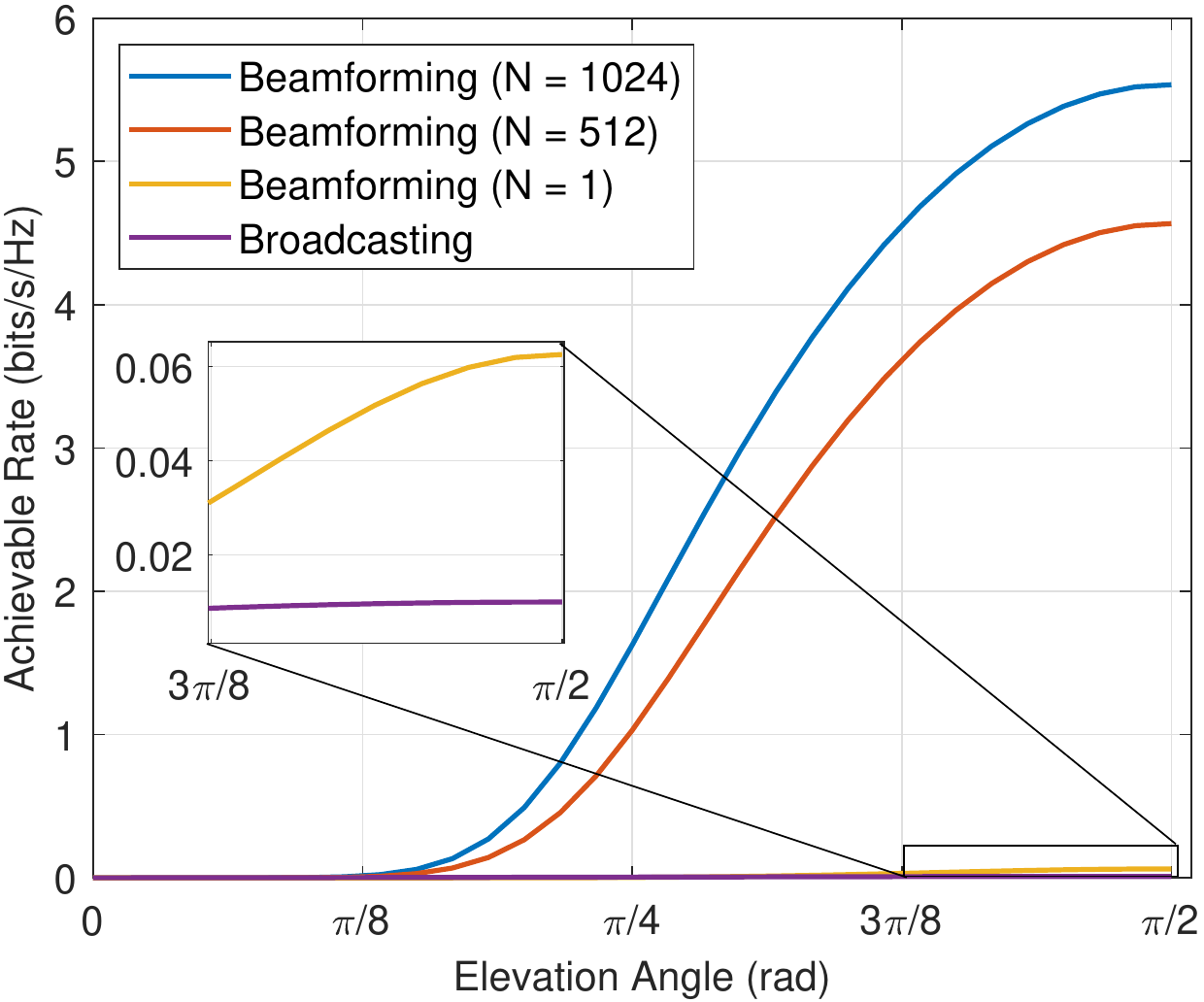}}
		\subfigure[]{
			\label{fig:capacity_vs_elevation_x_band_ul}
			\includegraphics[width=0.31\linewidth]{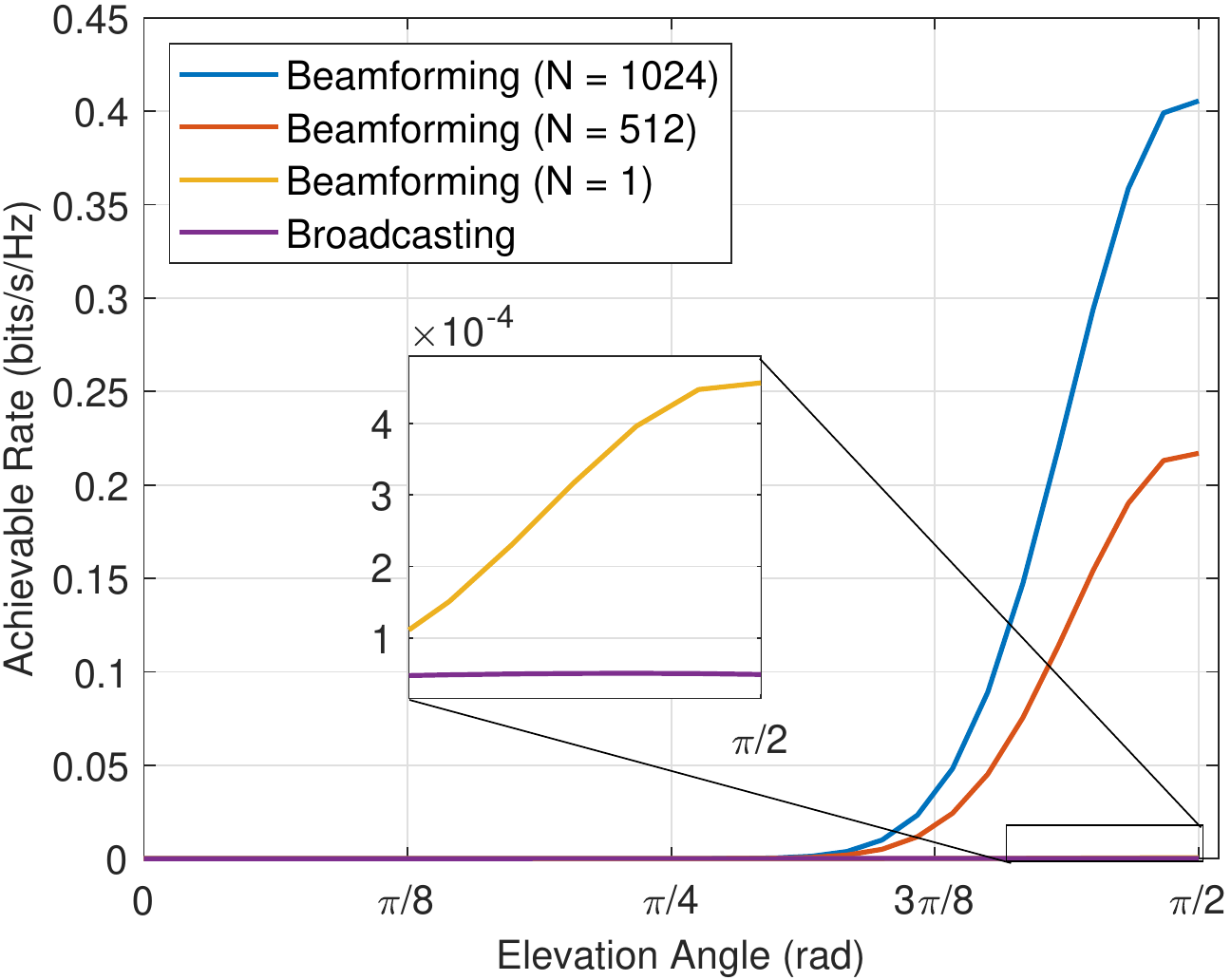}}
		\caption{Achievable uplink rate \textcolor{ black}{relative to} elevation angle $\varphi$ for (a) \textcolor{ black}{the} comparison between \textcolor{ black}{the case without RISs} and RIS-assisted satellites, (b) the various number of RIS elements in C-band, and (c) the various number of RIS elements in X-band. It should be noted that the transmit power is $1$ W.}
		\label{fig:ul_results_elevation}
	\end{figure*}
	
	In this section, we analyze the uplink capacity of satellite IoT networks in three cases: \textcolor{ black}{(i)} without RIS, \textcolor{ black}{(ii)} RIS broadcasting, and \textcolor{ black}{(iii)} RIS beamforming. \textcolor{ black}{As with the analyses of the downlink in the previous section,} we \textcolor{ black}{first consider} the advantage of \textcolor{ black}{using} RISs in satellites for IoT networks. \FGR{fig:capacity_vs_pt_with_without_ris_c_band_ul} shows that \ac{RIS} beamforming can provide \textcolor{ black}{an uplink rate that is $10^5$ times higher than the case without RIS.} This result indicates that the battery and lifetime of battery-limited \ac{IoT} devices can be significantly increased with RIS-assisted satellites. Moreover, \FGR{fig:capacity_vs_pt_c_band_ul} shows that it is possible to obtain $3$ dB more transmit power gain by doubling the number of RIS elements (from $512$ to $1024$) for the same achievable data rate value. \textcolor{ black}{As we can see in} \FGR{fig:capacity_vs_pt_x_band_ul} that the uplink data rate decreases significantly \textcolor{ black}{when} the operating frequency increases. As the number of RIS elements increases, the rate of decrease in the achievable rate due to the increase in frequency degrades. Although increasing the operating frequency reduces the achievable data rate performance, the higher frequency can be employed for small \ac{IoT} \textcolor{ black}{devices}, which cannot accommodate larger antennas. \FGR{fig:capacity_vs_elevation_with_without_ris_c_band_ul} shows that the RIS-beamforming scheme can \textcolor{ black}{offer} higher performance over a wider range of elevation angles than RIS broadcasting and non-RIS satellites. \textcolor{ black}{As we can see in} \FGR{fig:capacity_vs_elevation_c_band_ul}, increasing the number of RIS elements marginally improves the operation range in terms of elevation angle. \textcolor{ black}{The} higher frequency limits the communications in a narrow elevation angle range as shown in \FGR{fig:capacity_vs_elevation_x_band_ul}, which means that the communication duration in the period of a satellite is very short. \textcolor{ black}{Hence, using higher frequency} may cause many handovers between satellites in a short while.
	
	\textcolor{ black}{While} the computational capacities and batteries of IoT \textcolor{ black}{devices} are \textcolor{ black}{limited}, \textcolor{ black}{using \acp{RIS} can extend battery life since RISs can decrease the computational cost and transmit power for \ac{IoT} devices.} RIS-assisted satellite systems have \textcolor{ black}{a} two-fold advantages for IoT networks. First, RISs can provide \textcolor{ black}{higher} capacity \textcolor{ black}{with} lower transmission power. \textcolor{ black}{Second, they} enable complex operations to be performed in an external environment (i.e. propagation medium) rather than on the devices.

	\section{Open Issues and Research Directions}\label{sec:open_issues}
	
	\textcolor{ black}{Although} there are \textcolor{ black}{many} open issues for \ac{RIS}-assisted wireless communications, we \textcolor{ black}{focus here on} three of the most crucial issues for \ac{RIS}-assisted satellites for \ac{IoT} networks. \textcolor{ black}{These} are the channel estimation \ac{RIS} deployment on satellites, and simultaneous wireless information and power transfer (SWIPT), with a focus on \textcolor{ black}{the} physical layer. 
	
	\subsection{Channel Estimation}
	
	Since \acp{RIS} can change the amplitude and/or phase of the incident electromagnetic wave, they can largely eliminate the randomness of the propagation medium. However, \textcolor{ black}{\acp{RIS} must have} high-quality channel state knowledge to \textcolor{ black}{achieve} maximum performance. Channel estimation thus plays a critical role in satellite communications for \ac{IoT} networks \textcolor{ black}{in satisfying} the desired \ac{QoS}. Deep learning may be used for this purpose in order to achieve high efficiency under convincing channel conditions in channel estimation. \textcolor{ black}{In a recent work on channel estimation in \ac{RIS}-assisted backhaul communications, we proposed a channel estimation framework based on \acp{GAT} in~\cite{tekbiyik2020channel}. By considering unseen nodes into consideration, the \ac{GAT} can reduce computational complexity and improve learning performance. In the training process, the obtained signal samples obtained along with known pilot samples were assigned to the nodes and vertices of graphs, respectively. Since \ac{IoT} devices lack the hardware required for training and using the deep learning model, it is reasonable to deploy the proposed \ac{GAT} model to \ac{LEO} satellites rather than \ac{IoT} devices.} By utilizing channel reciprocity, the channel coefficients can be obtained through the uplink pilot signaling. Without the need for downlink pilot signaling, phase adjustment can be performed by \textcolor{ black}{the} \ac{RIS} by utilizing the coefficients found for the uplink channel. However, \textcolor{ black}{it should be noted that} this channel estimation cannot be employed in frequency division duplex communications for uplink and downlink. When using time division duplex, the signal processing can be performed through the propagation environment and \ac{RIS} rather than \ac{IoT} devices, thus \textcolor{ black}{extending} battery life.
	
	\vspace{-0.2cm}
	\subsection{RIS Fabrication and Deployment}
	
	\textcolor{ black}{In this study, we have assumed that \acp{RIS} are ideal devices that can support lossless reflection and continuous phase shifting. Although this assumption would not entirely hold good in practical application, it is important in terms of showing the upper performance of the system. Our recent study~\cite{tekbiyik2021graph} investigated the performance of channel estimation and phase configuration for nonideal \acp{RIS} in satellite-IoT systems. It was shown that \acp{RIS} with a sufficient number of discrete phase levels (e.g. 3-bits) can perform almost as well as continuous phase \acp{RIS}.} Also, environmental conditions must be considered in the design of \acp{RIS}. \acp{RIS} that can be resilient to temperature variations between day and night should be developed. Although space may \textcolor{ black}{indeed be a vacuum, it is also filled with plasma and other particles emitted from the Sun}. The electronic \textcolor{ black}{components} of the communication unit may be affected by charged particles encountered by spacecraft in the Van Allen Belts. As a result, it is crucial to consider radiation-resistant RISs and \textcolor{ black}{any other innovations that would allow RISs to remain operational under such conditions}.
	
	How to implement RISs in LEO satellites is undeniably \textcolor{ black}{the first and} the most fundamental question. \textcolor{ black}{Clearly this is a significant design challenge that needs to be addressed. In particular}, depending on variations \textcolor{ black}{in the positioning and elevation angle of the RIS units}, the device \textcolor{ black}{may} shadow any or all of the meta-atoms. \textcolor{ black}{Moreover, as discussed above, the position of \ac{RIS} units with respect to the transmit antenna (i.e., $\theta_{n}^{tx} = \theta_{n}^{t} = 0$) is important to maximizing the performance of the system. Thus, RIS deployment should be carefully considered.} Based on the development of their structures, the authors envisage that conformal metasurfaces can be used in the coating of objects with irregular surfaces and arbitrary shapes. It is worth \textcolor{ black}{recalling} that satellite systems have already utilized reflectarray antennas to reflect the incident beam with a constant phase~\cite{huang2005reflectarray}. These reflectarrays can be replaced by \acp{RIS}. \textcolor{ black}{This would allow satellites to reflect or scatter incident waves} with various phase configurations. Interdisciplinary studies can make novel RIS designs possible. For example, it may be possible to place RIS \textcolor{ black}{units over} a wide area by covering the lower faces of the satellite solar panels. Moreover, the antenna subsystems currently on satellites might be replaced by RISs.
	
	\subsection{Simultaneous Wireless Information and Power Transfer}
	As IoT nodes are power-limited devices, energy harvesting can be considered in \textcolor{ black}{a} SWIPT framework. RIS-aided SWIPT has \textcolor{ black}{already been proposed for achieving} significant performance gains in energy harvesting~\cite{wu2019weighted}. SWIPT can \textcolor{ black}{also} be a solution to \textcolor{ black}{keeping} power-constrained IoT systems running for a long time. But, its applications in satellite-IoT systems are lacking in the literature.
	Considering the power transfer capacity of space solar satellites with microwave waves~\cite{bergsrud2016rectenna}, the joint utilization of RISs and rectenna arrays can further improve the energy-efficiency of IoT networks. \textcolor{ black}{To put it briefly,}, RIS-assisted satellites for SWIPT in IoT networks deserve more attention in the future.

	\section{Concluding Remarks}\label{sec:conclusion}
	
	Ubiquitous connectivity and user-centric communications are the strict requirements of 6G networks. As the number of connected devices and \ac{IoT} networks increases, LEO satellite networks \textcolor{ black}{will gain increasing} as the backhaul/fronthaul connectivity solutions \textcolor{ black}{for} densely deployed IoT sensors. However, current systems either \textcolor{ black}{suffer from} high path loss or require steerable antennas \textcolor{ black}{on} IoT devices. \textcolor{ black}{Given these drawbacks,} we \textcolor{ black}{introduced an RIS-assisted LEO satellite framework} for energy-efficient IoT networks in this study. The motivation behind the RIS communications in satellite-IoT \textcolor{ black}{was} given using extensive numerical results throughout the study. The potential gain in terms of the transmission powers of lightweight IoT nodes \textcolor{ black}{were} quantified. Furthermore, open issues regarding the proposed system model \textcolor{ black}{were} also discussed.

	\balance

	\bibliographystyle{IEEEtran}
	\bibliography{ieee_iotj_ris_black}
	
	\balance
\end{document}